\documentclass[epj]{svjour}
\usepackage{graphics}
\usepackage[T1]{fontenc}
\usepackage[latin9]{inputenc}
\usepackage{amsmath}
\usepackage{amssymb}
\usepackage{graphicx}
\begin{document}
\title{Growth of a tree with allocations rules: Part 2 Dynamics}
\author{Olivier Bui\inst{1} \and Xavier Leoncini\inst{1}
\thanks{\emph{Present address:} Insert the address here if needed}%
}                     
%
%
\institute{Aix Marseille Univ, Université de Toulon, CNRS, CPT, Marseille, France}
\date{Received: date / Revised version: date}
%
\abstract{
Following up on a previous work\cite{Bui2019} we examine a model
of transportation network in some source-sink flow paradigm subjected
to growth and resource allocation. The model is inspired from plants,
and we add rules and factors that are analogous to what plants are
subjected to. We study how different resource allocation schemes affect
the tree and how the schemes interact with additional factors such
as embedding the network into a 3D space and applying gravity or shading.
The different outcomes are discussed.
} 
\maketitle

\section{Introduction}

River basins, roads, public transit network, water supply
network, electrical grid and vascular systems in plants and animals
have in common that they are systems transporting objects or substances
throughout a complex network. Each of these ``transportation networks''
have their own particularities which have been documented in specific
literatures such as urban transport\cite{FARAHANI2013}, river basins
\cite{Rodriguez92}, vascular system in animals\cite{Nguyen2006}
and plants\cite{west1999} etc. There is also more general literatures
whereby unspecified transportation networks are studied through an
abstract and mathematical lens\cite{Fulkerson1977,banavar1999size,Durand2007,Bohn2007,Carmi2008,Hougaard2009,Corson2010}.
In this case, the study revolves around an optimization, maximization
or minimization problem. Thus the focus is usually on the topology
of the network, either to construct an optimal network or to find
an optimal path, though we also do have work focusing on the geometry
of the pipelines\cite{Durand2006}, but the commonality is that the
substance being transported rarely plays an active role. In a foregoing
paper\cite{Bui2019}, we decided to change this paradigm by allowing
the substance to interact with the network itself.

The principle is simple: our transportation network is now a dynamical
system capable of growing or decaying, i.e. nodes are created and
deleted, and the driver of the growth is the transported substance
itself. It is treated as a vital resource that needs to be consumed
in order to sustain a ``node'' alive and can be also used to create
new nodes. Another particularity of the model is the addition of aspects
related to locality: the network does not follow an established plan
for what it should grow into, instead the network is built as the
result of nodes acting more or less individually to form the global
structure. This approach is largely inspired by vascular networks
in biology such as perennial plants: they do not have a heart or a
brain that could single-handedly control the growth or the distribution
of resource throughout the plants and yet they end up with consistent
shapes. Trees could be an example of self-organization arising from
local rules and interactions\cite{SACHS2004197}. This is why the
previous paper\cite{Bui2019} as well as the current one will draw
its kinematic and dynamic rules as well as its terminology from plants:
the network we are studying are called ``trees'', are represented
by a connected acyclic graph with the initial node called a ``root''.
At the extremities, the nodes are ``leaves'' and would continuously
produce the vital resource which would be transported throughout the
tree. Such model is supposed to be analogous to phloem transport,
the phloem being the part of the tree that handles the transport of
the sugar photosynthesized at the leaves\cite{DeSchepper2013,Jensen2016}.
Though the relation between the models introduced in the present paper
and real plants should not be taken beyond an analogy, it can be noted
that the literature on plants has a wealth of quite simplistic theoretical
or numerical models to explain specific aspects of plants\cite{Niklas1994a,Eloy2017b,Duchemin2018}.

The organization of the paper is as follows, we start with a description
of the model as well as the results and conclusions found in \cite{Bui2019}
relevant for the present paper in Sect.~\ref{sec:A-bio-inspired-model}.
This is followed by Sect.~\ref{sec:Example-of-implementation-1}
where we see how the systems react to some control and try to shape
the tree into diverse forms. In Sect.~\ref{sec:A-static-description}
we perform some analytical calculations and, try to explain and summarize
the main takeaway from the model before concluding in Sect.~\ref{sec:Conclusion}.

\section{Description of the model\label{sec:A-bio-inspired-model}}

In this Section, we will describe the model as well as summarize the
results found in the foregoing paper\cite{Bui2019}. The description
of the model is divided into two Subsections. The first one referred
as ``Kinematics'' details the basic rules of the model. The second
Subsection specifies how the branches interact with one another.

\subsection{Kinematics\label{subsec:Kinematics}}

In this Subsection, we describe the basic rules of the model and summarize
some results about it.

\subsubsection{Model\label{subsec:Model}}

\begin{figure}
\includegraphics[width=130bp]{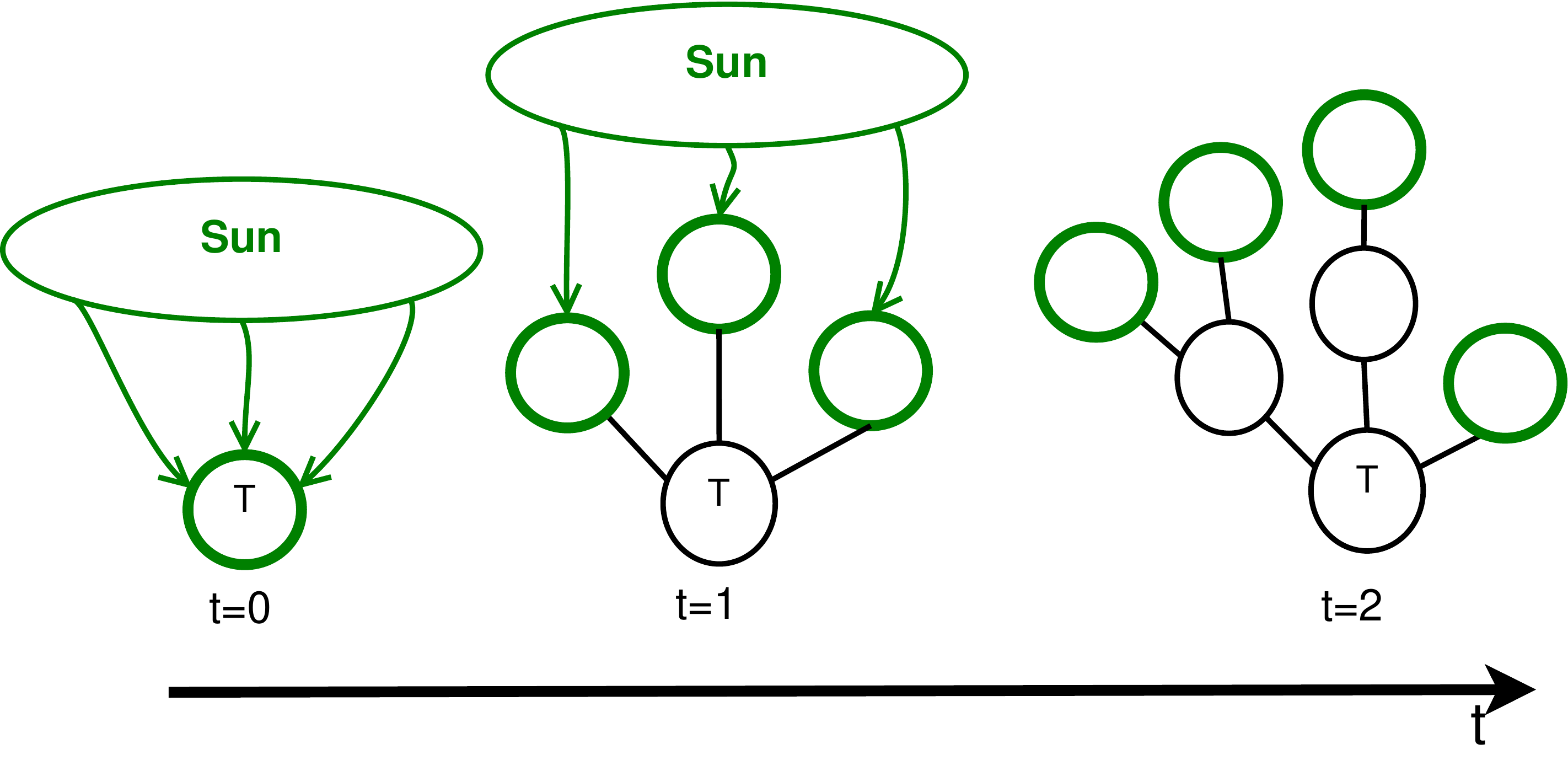}\includegraphics[width=120bp]{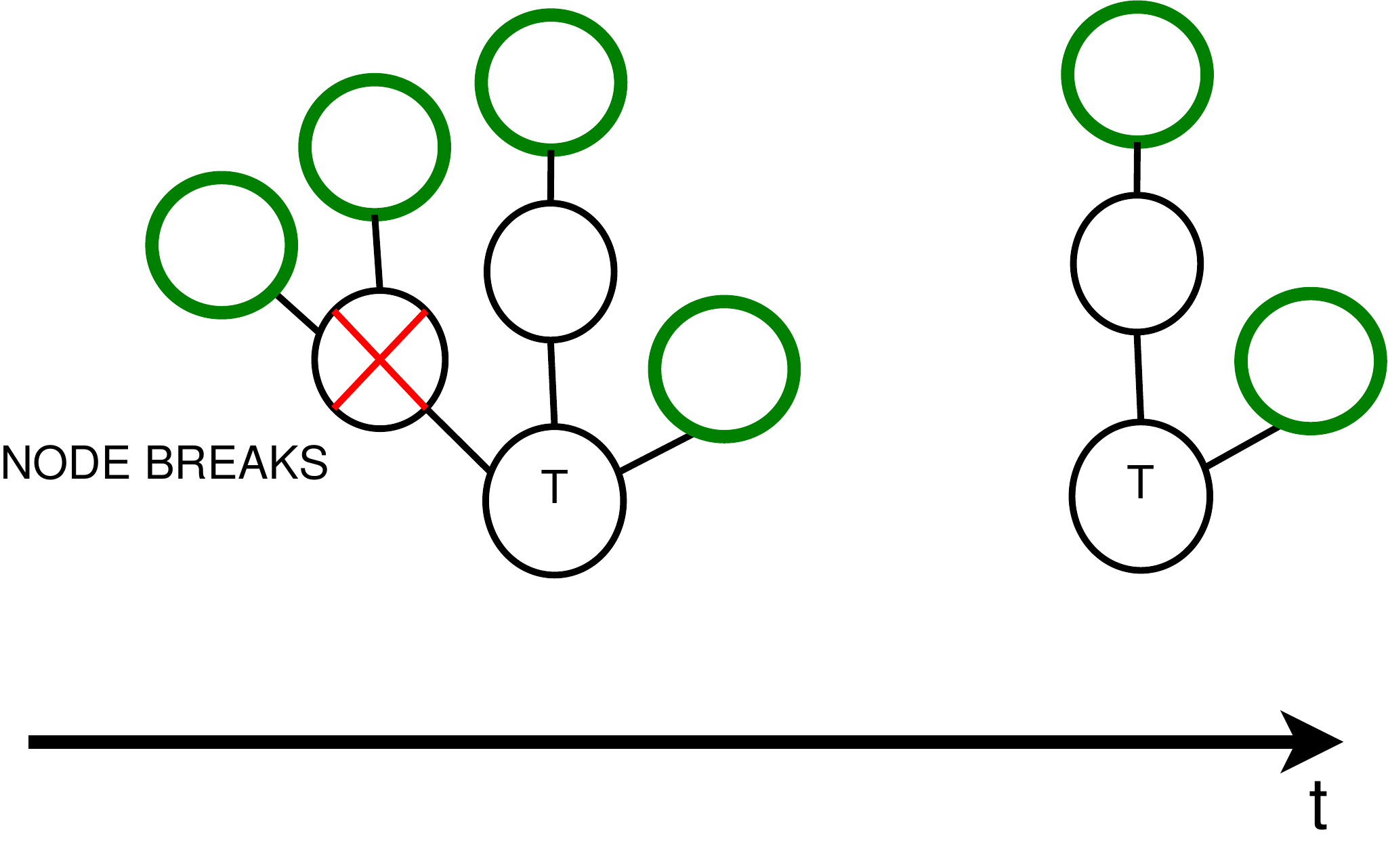}\caption{\label{fig:1}Representation of what the model is about: a growing
tree taking ``resource'' from its leaves, using the resource to
grow and feed its different parts with the consequence of death for
the malnourished parts.}
\end{figure}
Let us summarize the basic model we used in \cite{Bui2019}.\textit{
Trees} are modeled as graphs, where the nodes are called \textit{branches}.
We will use terminology such as \textit{child} and \textit{parent}
to describe the relation between branches. The tree initially starts
as a graph with only a single node. This node will be called the ``root''.
The graph will be able to grow by producing more branches through
a few simple rules which are essentially local, while the last one
introduces non locality:
\begin{itemize}
\item Time is discrete. Each branch with no child, we will refer to as extremity,
produces some quantity $p_{0}$ of resource per unit of time.
\item The resource is used by extremities to create children, and only extremities
can create new children, at a cost of some $C_{r}$ resource per child.
A branch creating children ceases to be an extremity. A branch can
only have $N_{max}$ children, at most.
\item Each branch has a certain width. The branch tries to increase its
``width'' so that it is ``equal'' to the number of extremities
that are its \textit{descendants} (children or grandchildren etc.).
This ``width'' will be called \textit{volume} of a branch $V$ This
appellation is still consistent with our plant analogy as what we
call branches would actually be sections of branches of the same length,
thus making a branch cross-section area ``equal'' to its volume.
As for the rule that $V$ must at least be equal to the extremities,
we will call it ``Leonardo's rule'' in reference of the real rule\cite{leonardo_dav,Minamino2014}.
\item Each branch must consume resource to grow in volume with a cost of
$C_{m}$ multiplied by the quantity of volume created. Furthermore
they must pay a maintenance cost of $m_{0}\times V{}^{\alpha}$ per
unit of time. $\alpha=1/2$ is the value used by default.
\item If a branch \textit{dies} (could not pay its maintenance cost) then
all its descendants will die.
\end{itemize}
\begin{figure}
\includegraphics[width=100bp]{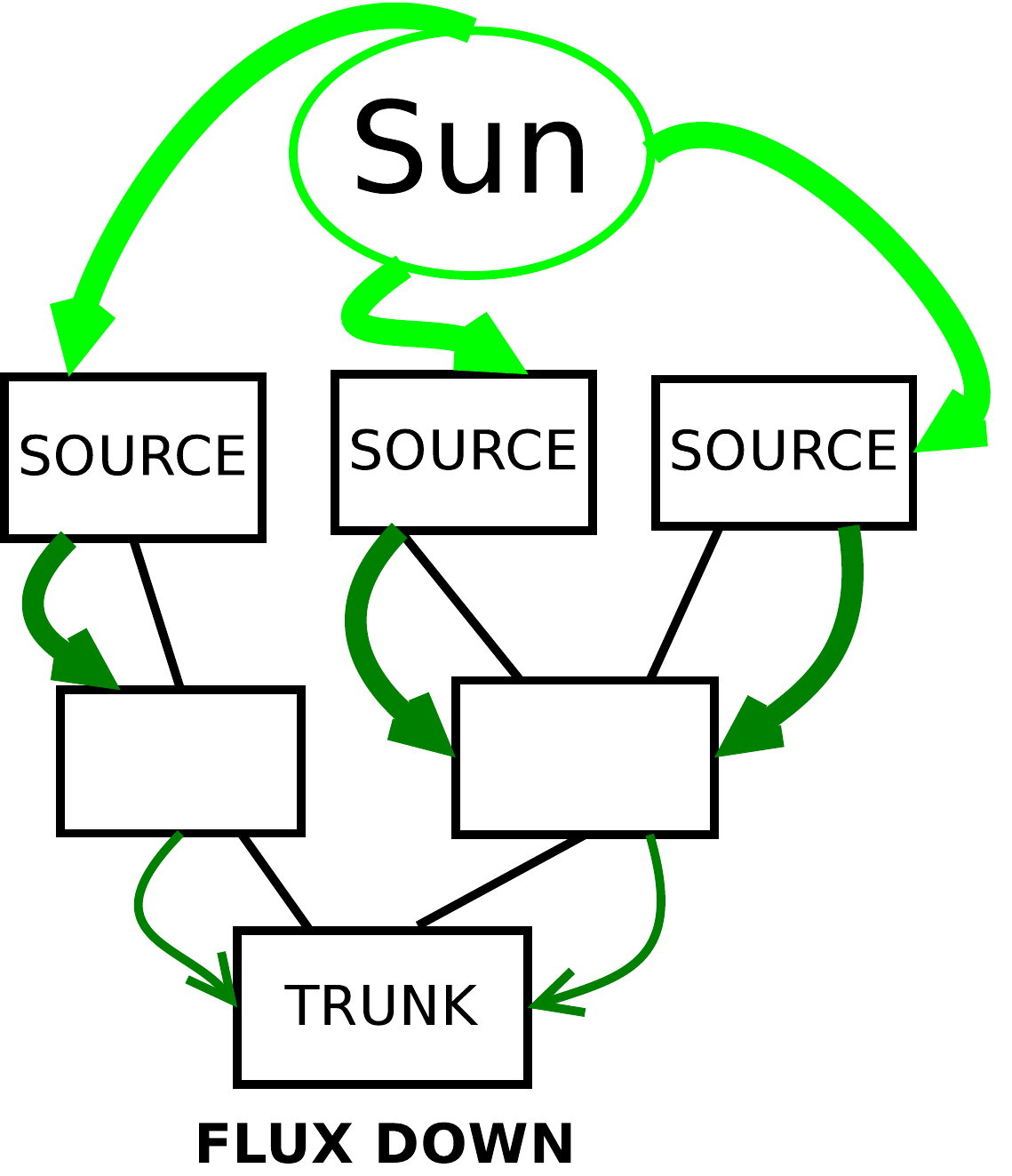}\includegraphics[width=100bp]{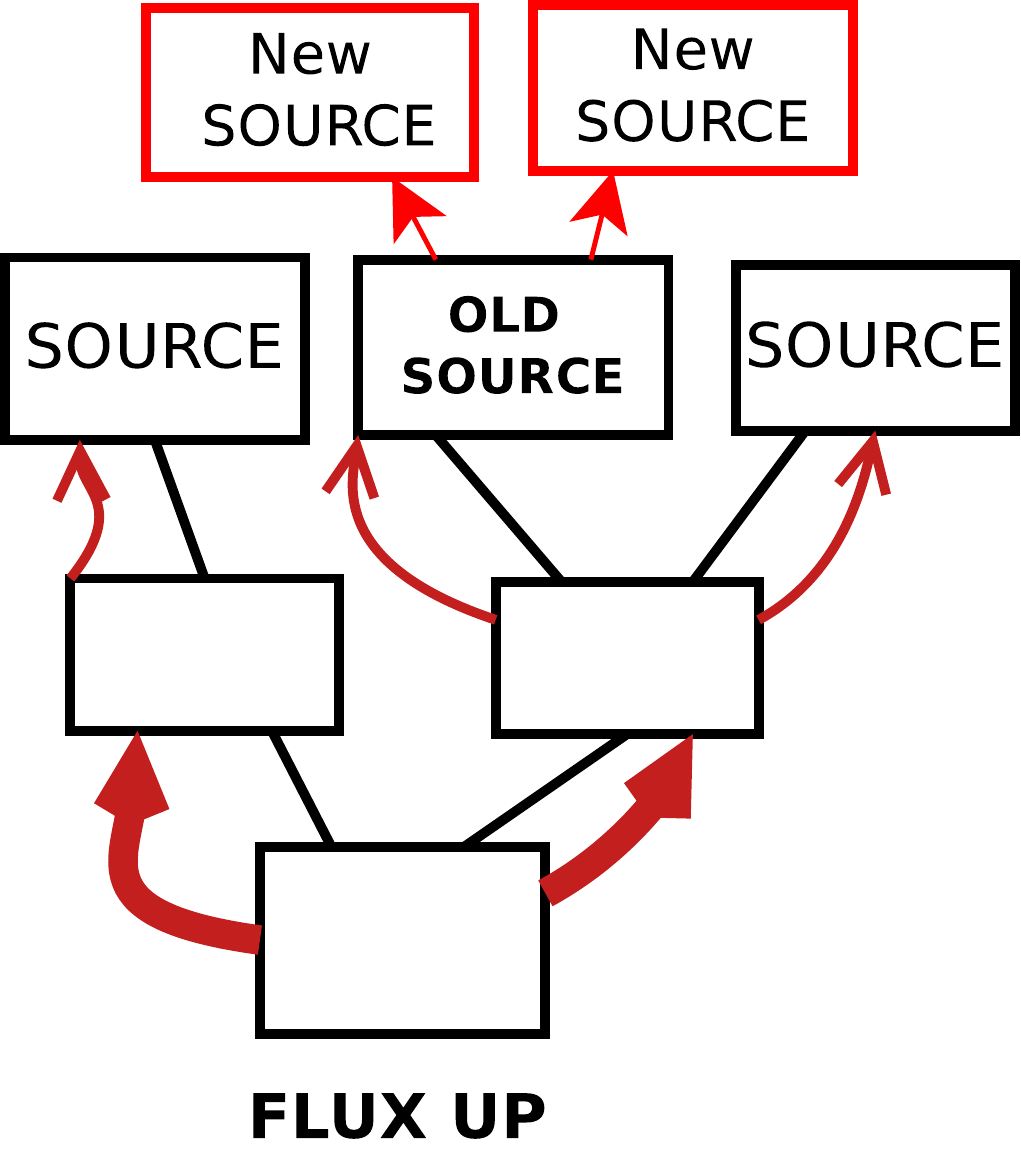}\caption{\label{fig:2}A representation of the ``flux down'' and ``flux
up'' kinematic described in Sect.~\ref{subsec:Model}. Leaves create
a quantity of resource, this resource flows down to the ``root''
of the network with each intermediate nodes saving in their memory
the amount of ``flux'' given by each of their child. Then the parents
starting from the ``root'' will share their resource with their
children according to factors such as the flux given by each child
previously.}
\end{figure}
With only extremities being able to produce resource but every branches
needing to use it for sustenance, we must add to the model a scheme
for allocating the resource throughout the tree (Fig.~\ref{fig:2}).
The scheme or kinematics is in two part, one called ``flux down''
whereby the extremities produce the resource then all the resource
flows down the tree to be gathered at the ``root''. A caveat is:
during the ``flux down'', the children will not transmit all the
flux down their parent, instead they will consume enough of it to
grow in volume $V$ according to the ``Leonardo's rule'' discussed
earlier and only thereafter transmit the flux down. Then we have a
``flux up'' where allocation choices are really made:
\begin{enumerate}
\item If the branch is the root it starts with all the resource gathered.
Otherwise it starts with the amount its parent has decided to send
(see point number 3).
\item Using the resource gathered, it pays the maintenance cost based on
the volume. The branch breaks otherwise.
\item The branch send all its remaining resource to its children. The share
given to each one depends on the amount the child gave back during
flux down or on other factors discussed later. The ``sharing'' scheme
dictates the dynamics of the tree. Then we go back to point number
1 above.
\item If the branch has no child (it is an extremity) then it uses the resource
to create children. There is a maximum number of children it can create.
Leftover are kept. During the next unit of time (we call it \textit{generation}),
it will be that leftover in addition to the resource produce from
photosynthesis that will flow down to the root.
\end{enumerate}

\subsubsection{Previous results\label{subsec:Previous-results}}

There are a few things we noted in the previous paper\cite{Bui2019}
that need to be reminded in the present paper which we will do in
this Section. First, the values of the production parameter $p_{0}$,
maintenance $m_{0}$, volume growth $C_{m}$ and cost of branch creation
$C_{r}$ are not important: it is actually the ratio between those
parameters that are the important factors. More specifically we can
reduce the set of parameters into $p_{0}/C_{r}$, $m_{0}/C_{r}$ and
$C_{m}/C_{r}$ or simply set $C_{r}=1$. For the rest of the paper,
when we talk about $p_{0}$, $m_{0}$ and $C_{m}$, we will actually
refer to $p_{0}/C_{r}$, $m_{0}/C_{r}$ and $C_{m}/C_{r}$ respectively.

On the other hand, the exponent $\alpha$ appearing in the maintenance
formula ($m_{0}\times V^{\alpha}$) strongly determines how much the
tree can grow and how long it survives. Theoretical calculations can
also determine (assuming $N_{max}<\infty$ which would only be false
if our tree lives in an infinite dimension space) whether there exists
tree of any given heights with a positive ``balance'', which we
define as the total production of the tree minus its total maintenance
cost. One result we established was that at $\alpha=1$ or higher,
we can always find some heights beyond which no tree will have a positive
balance. Only if $\alpha<1$ can we find arbitrarily tall tree, thus
opening the possibility of an infinitely growing tree. However, because
of the cost $C_{m}$ for branches to grow in volume, we also established
that an infinitely growing tree would need to exponentially slower
its growth rate as the tree grows taller. Other constraints are needed
on parameters such as $p_{0}$, $m_{0}$ and $N_{max}$ to allow the
existence of infinitely growing trees, more specifically we need $p_{0}/m_{0}\geq\dfrac{N_{max}^{1-\alpha}}{N_{max}^{1-\alpha}-1}$
to be true.

The takeaway is reducing $C_{m}$ to 0, $\alpha<1$ and a proper value
for $p_{0}/m_{0}$ allow the simulation of infinitely growing tree
at a constant rate. It must be noted that the results on infinitely
growing tree is true even if we do not use the kinematics driven by
the ``flux down'' - ''flux up'' we described in Sect.~\ref{subsec:Model},
instead they derive solely from the five ``simple rules'' listed
prior the ``flux'' part and as such are truths that go beyond the
particular kinematics presented in the paper. Nevertheless, to perform
simulations we need a kinematics and we use the ``flux down and up''
one, setting $C_{m}=0$ and $0<\alpha<1$. When it comes to how parents
should split their resource during ``flux up'', we choose an equal
share to be given to each child. Details are presented in the previous
paper but the main takeaway is the system was quite ``uneventful''.
Each tree growth simulation could be fully characterized by a set
of integers which itself could be summarized by three numbers, and
when we slowly vary the initial parameters it had small and predictable
effects on the tree. In other words, there was no trace of any kind
of ``chaotic'' or complex behavior.

\subsection{Dynamics\label{subsec:Dynamics:-putting-the}}

In this Subsection, we specify how the branches interact with one
another. We begin by introducing the resource redistribution schemes,
detailing what we could expect them to do and performing some simulations
to look into it. Finally, we introduce another way for the branches
to interact: an embedding into a three-dimensional space.

\subsubsection{Allocation and distribution of resource\label{subsec:Allocation-and-distribution}}

The form of the tree will be shaped by how the resource is distributed
across the tree. More precisely during the ``flux up'' phase when
each parent shares its resource among its children, the different
proportion given to each child will drive the shape of the tree. Sect.~\ref{subsec:Previous-results}
presented both general results as well as results specific to some
simulations. These simulations had the parents sharing equally among
its children its resource during ``flux up'', and it resulted in
uneventful results thus incentivizing us to use more complex redistribution
schemes. In the present paper, we study two types of redistribution
schemes: a first scheme where the parent shares a higher proportion
of its flux to children that during ``flux down'' was the more productive
and a second scheme where the parent shares more to children with
higher maintenance costs. A more formal description of the first scheme
is as follows: for a parent of $N$ children, if $c_{i}$ is the amount
of resource the child $i\in\{1,...,N\}$ gave to its parent during
the previous ``flux down'' phase, then the parent will give $c_{i}^{a}/Z_{1}$
resource to the child $i$ where $Z_{1}=\sum_{i}^{N}c_{i}^{a}$ is
a normalization constant. $a$ is a positive parameter and determines
the degree by which parents favor the more productive children. For
the second scheme, if we note $m_{i}$ the maintenance cost of the
child $i$ then the parent will distribute $m_{i}^{b}/Z_{2}$ resource
to the child $i$. Analogous to $a$, $b$ is a positive parameter.
The first redistribution scheme will be called the ``reward'' scheme
where as the second one will be the ``maintenance'' scheme. The
two schemes are summarized in Eq.~\ref{eq:3}. We call $F_{i}$ the
proportion of flux a parent will give to its child $i$ and the values
of $a$ and $b$ determine which scheme we use. The ``reward'' scheme
is obtained when $a>0$ and $b=0$ and the second one is obtained
when $b>0$ and $a=0$. If $Z=0$, we set $F_{i}=0$ for all $i$.
Unless explicitly specified, we will not mix the two schemes, as such
whenever $a>0$ it is implied $b=0$ and vice versa. 
\begin{equation}
F_{i}=\frac{c_{i}^{a}m_{i}^{b}}{Z}\textrm{ where }Z=\sum_{i}^{N}c_{i}^{a}m_{i}^{b}\label{eq:3}
\end{equation}

To get a grasp on what these schemes are supposed to do: let us assume
we use the ``reward'' distribution and look at a branch with 2 children.
Then if child 1 produced $x_{1}$ in resource while the child 2 produced
twice the amount which is $2x_{1}$, we would have $F_{2}=2^{a}F_{1}$.
Now let us look at what happen next generation. The side of tree emerging
from child 1 creates some new extremities thus now producing $x_{1}+x_{2}$
resource. But since child 2 previously received $2^{a}$ times more
resource than child 1, this side would, in the simplest case, have
produced $2^{a}$ times more new extremities resulting in just as
much more resource produced: $2^{a}(x_{1}+x_{2})$. So, for the next
``reward'' distribution, we get $F_{2}=(2^{a})^{a}F_{1}$. And we
can easily see the pattern for the following generations. This example
ignores plenty of factors, but, in this simple view, $a=1$ means
the proportions distributed between children would not change as time
passes, and the gap in resource allocation between branches will be
proportional to the total production. On the other hand, with $a>1$
would change the proportions in such a way that some branches may
get less and less resource even when the total production increases
($F_{2}/F_{1}$ grows exponentially in our example). What may prevent
this exponential inequality increase is the total maintenance costs
increasing as the tree ``grows'' which would impede or stop the
tree growth and thus stop the gap of widening.

The ``maintenance'' scheme (defined by $a=0,$ $b>0$ in Eq.~\ref{eq:3})
should not act much differently. Assuming the maintenance exponent
$\alpha$ is $1/2$ (the maintenance would scale like the square root
of the branch ``volume'') $b$ could yield the same trees as $a/2$
if we oversimplified the problem. Indeed the ``maintenance'' scheme
scales the portion $F_{i}$ with the branch maintenance cost, $m_{i}^{b}$,
but $m_{i}\propto V_{i}^{\alpha}$ and $V_{i}$ is just the number
of extremities attached to the branch, and if the contribution during
``flux down'', $c_{i}$, used for the ``reward'' scheme was proportional
to the number of extremities (the sources/leaves), the correspondence
``$b=2a$'' should appear.

\subsubsection{``Apical dominance'' in non-spatial tree\label{subsec:Non-spatial-tree}}

\begin{figure}
\includegraphics[width=200bp]{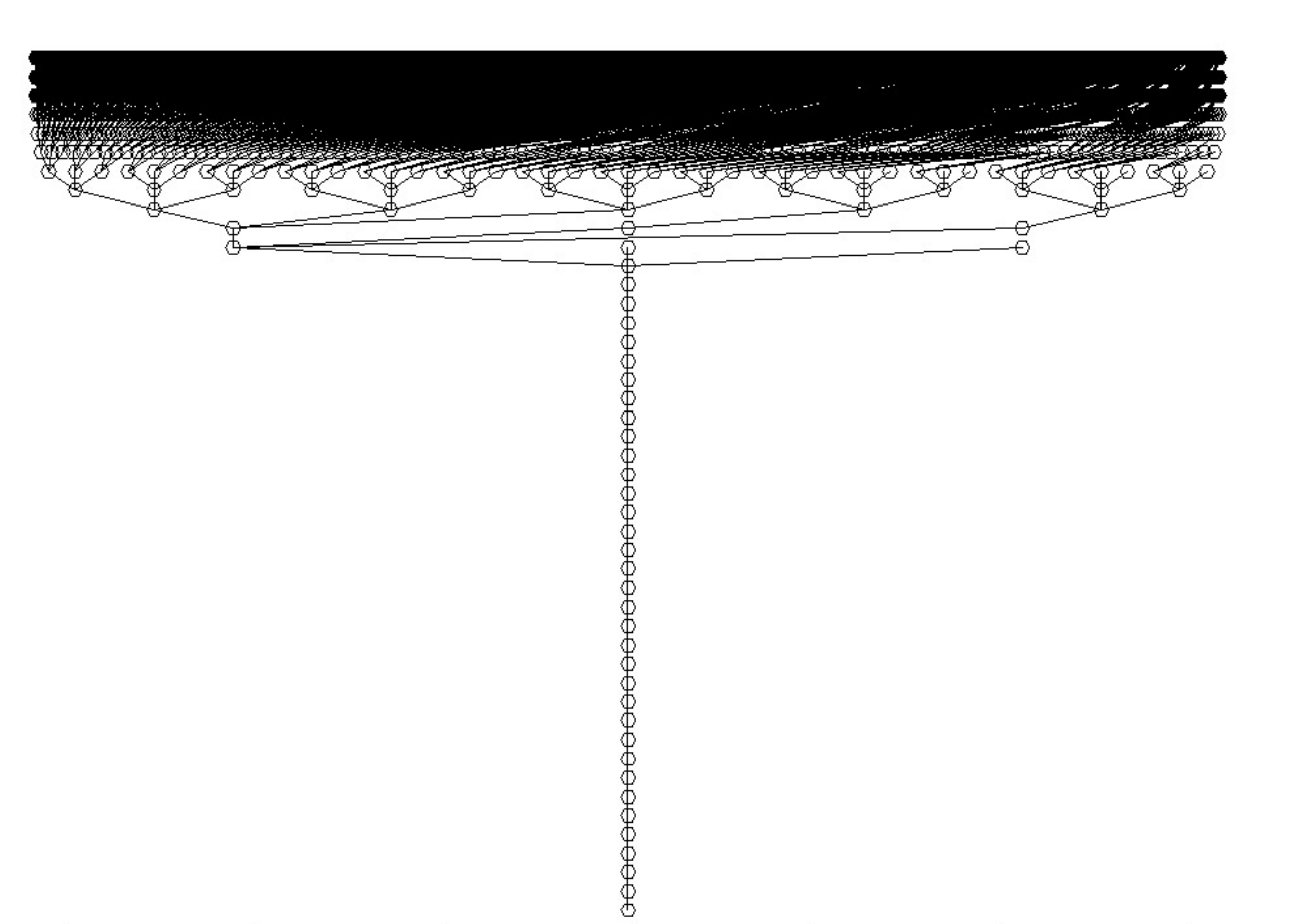}\caption{\label{fig:3}Tree simulated from the ``apical dominance'' rules
described in Sect.~\ref{subsec:Non-spatial-tree}. The parameters
used were $N_{max}=3$, $p_{0}=9.6$, $m_{0}=1$, $C_{m}=1$. The
``reward'' redistribution scheme is used with $a=1.5$ and the apical
dominance forces dominant parent to allocate at least 10\% of their
resource to their dominant child. Self-pruning end up appearing leading
to the formation of a trunk.}
\end{figure}
If we apply the ``reward'' or ``maintenance'' redistribution without
any additional factors, the tree will simply end up ``symmetric'',
indeed there would be no reason for Eq.~\ref{eq:3} to yield different
portions for each child $i$. To break this symmetry, we can give
a priority to some branches.

Here is the scheme proposed: during ``flux up'', when the initial
branch we called ``root'' has to distribute the flux to its children,
it will now have to allocate a fixed percentage, let us say $10\%$,
of the flux to the child labeled as its ``first child'' and only
then would the remaining $90\%$ be shared according to Eq.~\ref{eq:3}.
Then, the ``first child'' of the root will allocate $10\%$ to its
own ``first'' child before sharing the remaining $90\%$, the same
scheme is applied on this last ``first'' child etc. This scheme
is loosely analogous to ``apical dominance'' in trees whereby some
apical bud will grow more strongly\cite{cline1991apical}. Fig.~\ref{fig:3}
shows the result of the bias introduced in the ``apical'' scheme
when combined with the ``reward'' distribution scheme whereby more
resource are given to children that gave back the most. Self-pruning
has occurred and formed a ``trunk'' for $a=1.5$ which could be
justified by the explanation we gave in Sect.~\ref{subsec:Allocation-and-distribution}
about $a>1$ forcing branches producing less to progressively being
starved as the tree grows. Consistent with this explanation, we observe
no trunk for $a=1$, though for values very close to $1$ we may or
may not observe a trunk depending on the other parameters. It confirms
that the self-pruning is indeed caused by the redistribution scheme
and not simply due to the fact $10\%$ of the resource are allocated
to some branches. If the ``apical dominance'' scheme is only used
for the first 5 generations of the simulation (and we revert back
to a purely ``reward'' scheme afterward), then self-pruning does
not appear, at least when the $10\%$ number was picked, and instead
of a tree that would continue to grow in height the tree we get either
reaches some equilibrium or die.

So while a relatively small or moderate perturbation allows the ``reward''
redistribution scheme to drastically change the topology of the tree,
the perturbation should be ``sustained''. The system appears to
be resilient to short-term perturbation despite our talk in Sect.~\ref{subsec:Allocation-and-distribution}
about $F_{2}/F_{1}$ potentially increasing exponentially as time
passes. Predictably switching to ``maintenance'' scheme does not
yield $"b=a/2"$ since this equality relied on some relation between
maintenance and the amount of resource of a child would possess which
is explicitly broken because of the $10\%$ priority scheme.

Now let us look at a symmetry-breaking scheme that could yield the
$"b=a/2"$, and the scheme or ``perturbation'' should be sustained.

\subsubsection{Spatial embedding\label{subsec:Spatial-embedding}}

\begin{figure}
\includegraphics[width=180bp]{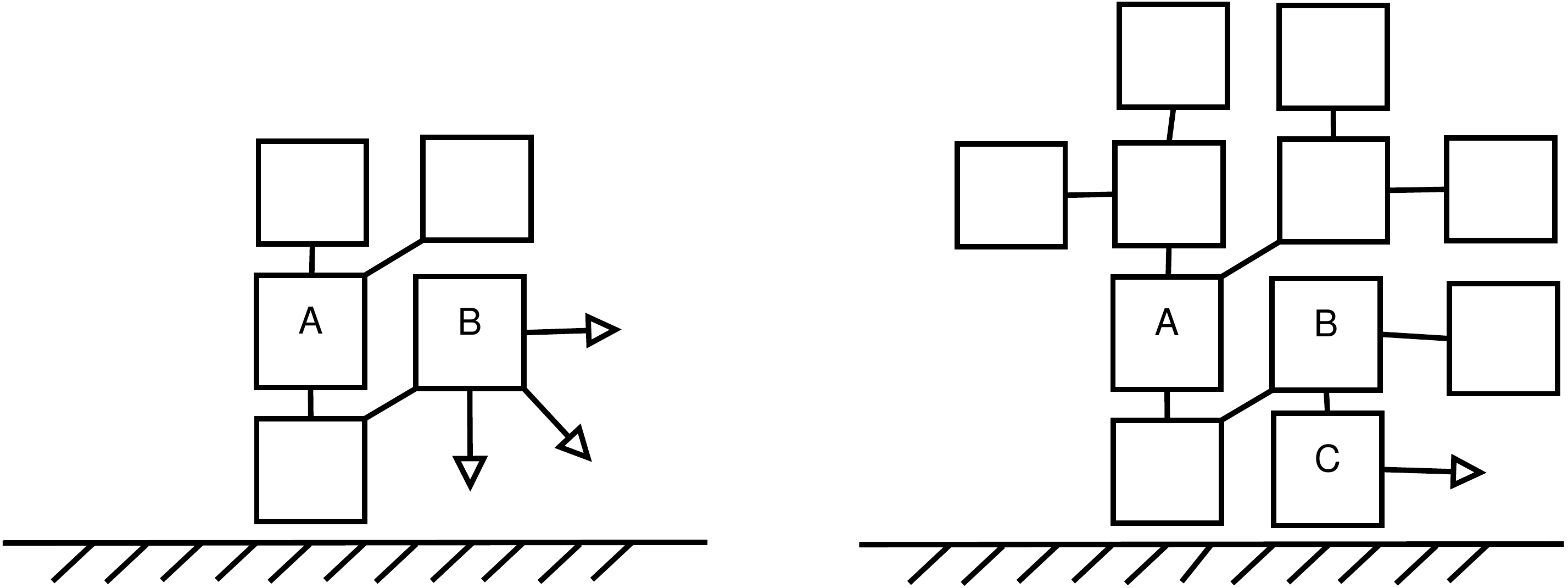}\caption{\label{fig:Fig4}Rough sketch of a 2D version of the spatial embedding
as detailed in Sect.~\ref{subsec:Spatial-embedding}: each node of
the network is now located in a square grid space. Two nodes can not
occupy the same coordinates and children and parents must in be adjacent
coordinates, with diagonal squares being considered adjacent.}
\end{figure}
We will now describe how the spatial component is added to the model.
\begin{itemize}
\item we get a three-dimensional space divided into cubes. Our tree is embedded
in this space and each branch occupies a single cube. Two branches
can not occupy the same position.
\item A future parent can only create children in an unoccupied adjacent
case, including diagonally adjacent cases. So a branch has 26 adjacent
cases. The ground which represents the plane right below the initial
branch, the ``root'', are considered occupied cases.
\item Branches do not decide the location of the children they will create
at the same time: we go through the future parents according to some
order. As such the ones at the bottom of this ordered list may find
themselves without enough space to create as many children they wanted
to have. This should create some asymmetries in the tree as exemplified
in a simplified two-dimensional version of our model in Fig.~\ref{fig:Fig4}.
\end{itemize}
The ordered list is not randomly generated at the start of each generation:
priorities in this ordered list are heritable meaning the descendants
of some branch $A$ that had priority over some branch $B$ during
last generation will have priority over branch $B$ or its descendants.
We can hope the fact branches do not decide their children's locations
simultaneously thus creating a hierarchy between parents is enough
to apply asymmetries as powerful as the one caused by the ``apical
dominance'' scheme.

On top of this simple spatial model, we will add different other factors
such as a gravitational factor and light interception and look at
how they interact with the ``reward'' and ``maintenance'' redistribution
schemes.

\section{Model exploration\label{sec:Example-of-implementation-1}}

Adding to the previously described model, we seek to control the growth
of the tree by adding more variables into the models. The added variables
are gravity and light interception.

\subsection{Children generated in random directions\label{subsec:Children-generated-in}}

\begin{figure}
\includegraphics[width=75bp]{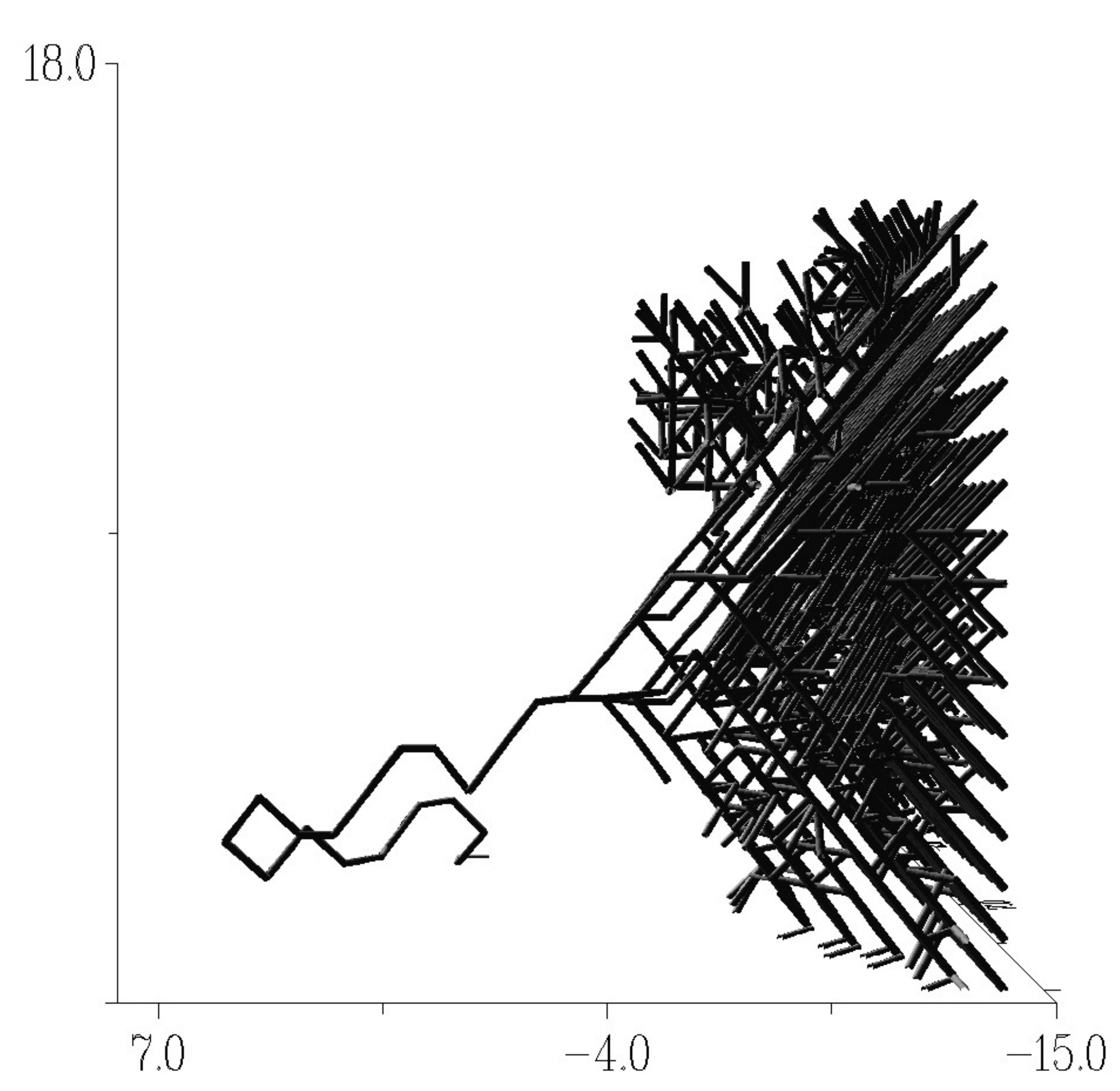}\includegraphics[width=75bp]{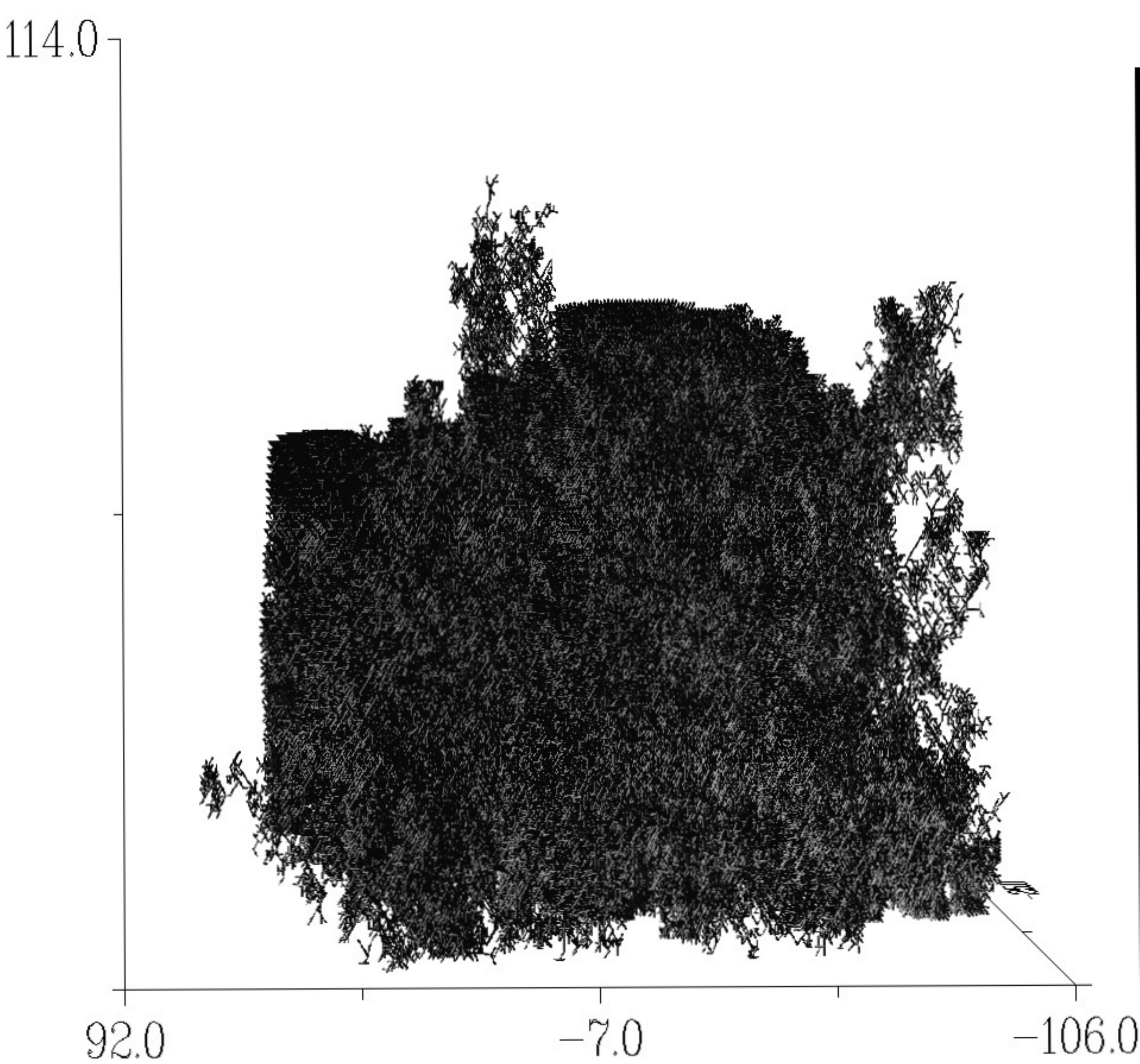}\includegraphics[width=75bp]{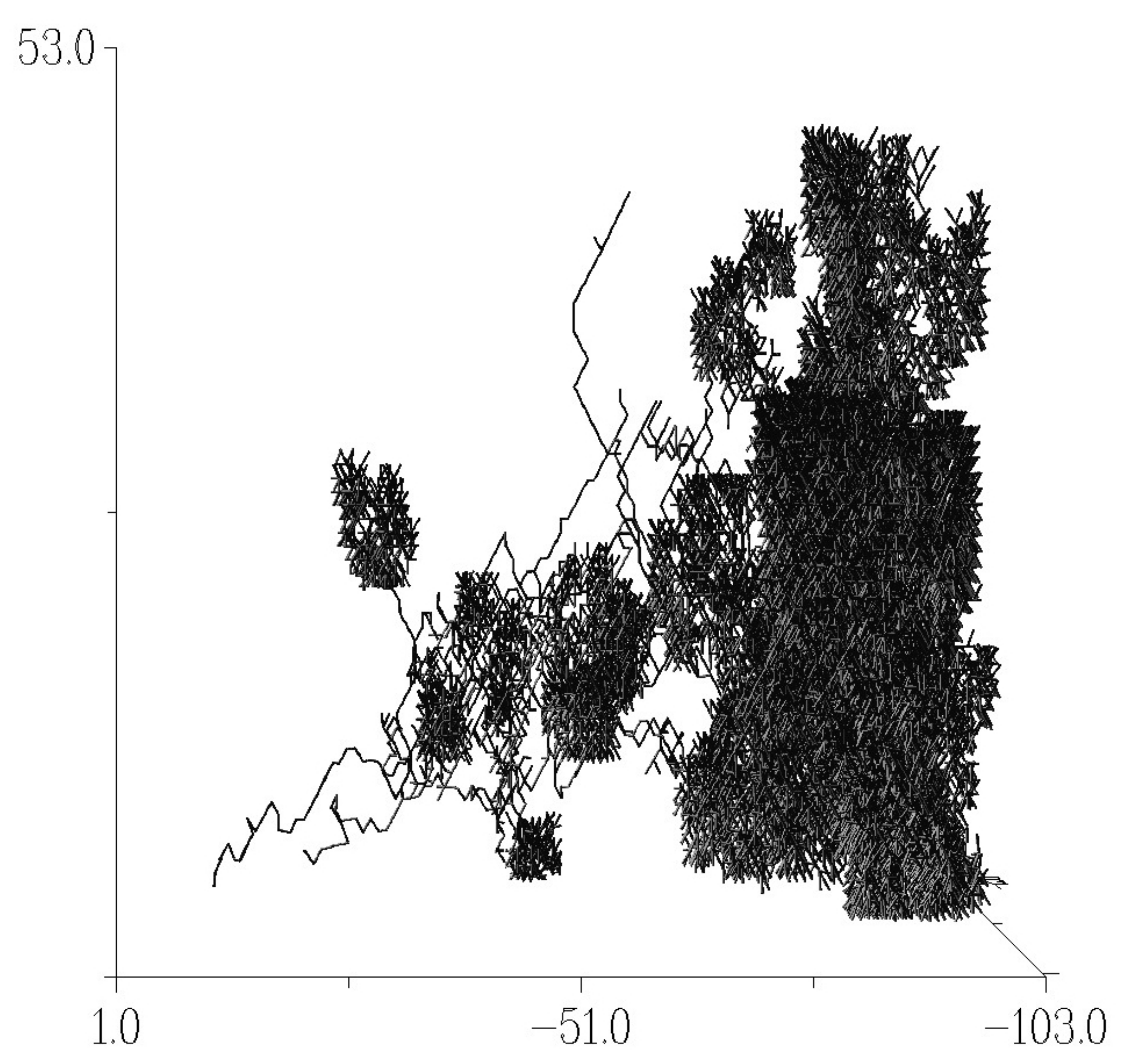}

\caption{\label{fig:Fig5} Simulation of trees embedded in space using $p_{0}=12$,
$m_{0}=1$, $C_{m}=1$, $N_{max}=3$ with children's location being
chosen at random. Each tree allocates its resource differently: The
leftmost tree uses the ``reward'' scheme at $a=2$, the middle tree
uses the ``maintenance'' scheme at $b=5$ and the last one is at
$b=6$.}
\end{figure}
 With the introduction of space in the model a new parameter emerges
which is the strategy a parent will have in order decide on the location
of its children. The simplest ``strategy'' is the random one: each
parent will create children in random available locations. We perform
simulations with the proposed random strategy combined with the redistribution
schemes described in Eq.~\ref{eq:3}: the tree always starts seemingly
expanding in all directions somewhat looking like the middle tree
of Fig.~\ref{fig:Fig5} then there is a divergence depending on the
redistribution scheme used. The ``reward'' distribution scheme,
whereby parents prioritize children that produced more resource, cause
the tree to self-prune depending on the coefficient value for $a$
used. For example, given the parameters in Fig.~\ref{fig:Fig5} and
$a>1.3$, we get the tree on the left: the tree self-pruned like in
Sect.~\ref{subsec:Non-spatial-tree} and created a ``trunk''. For
values of $a$ closer or equal to 1, the middle tree of Fig.~\ref{fig:Fig5}
is obtained: there is never enough self-pruning to form a substantial
``trunk'', the tree maintains a bush-like shape. Using the ``maintenance''
scheme $(b>0$ and $a=0)$, we also obtain a bush-like tree except
for very high values of $b$ (tree on the right in Fig.~\ref{fig:Fig5}):
a trunk is apparent but it branches out into multiple directions,
each ending with ``blobs'' of ``extremities'' that we will name
``clusters of leaves'' instead. This is largely different from the
tree on the left where all the leaves are concentrated together. The
main takeaway is that despite a lack of ``apical lead'' (see Sect.~\ref{subsec:Non-spatial-tree})
and an unsystematic way to decide the location of children we can
still end up with self-pruning. A more in-depth discussion is done
at Sect.~\ref{sec:A-static-description}.

\subsection{Non random directions for children generations\label{subsec:Non-random-directions}}

\begin{figure}
\includegraphics[width=110bp]{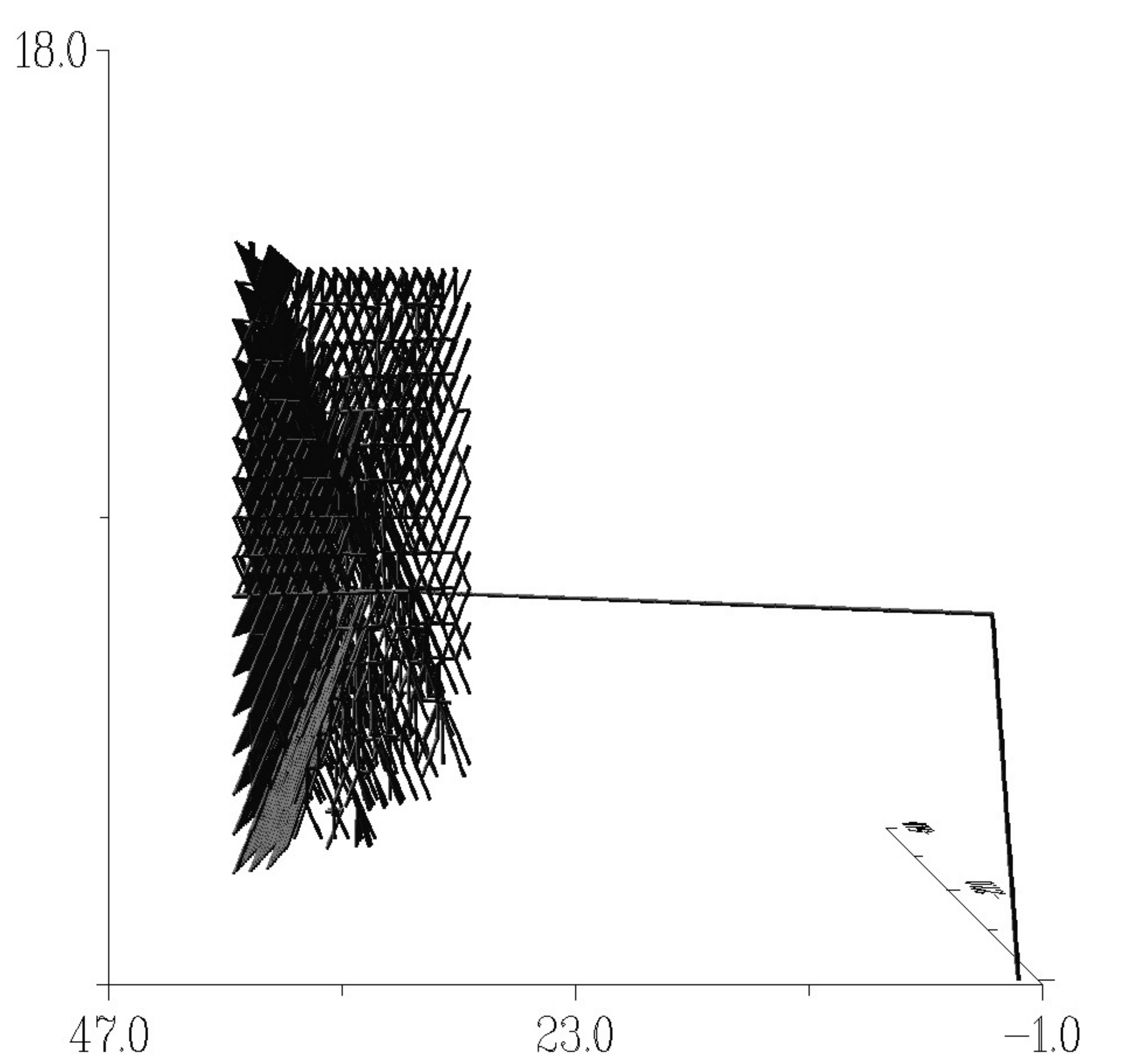}\includegraphics[width=110bp]{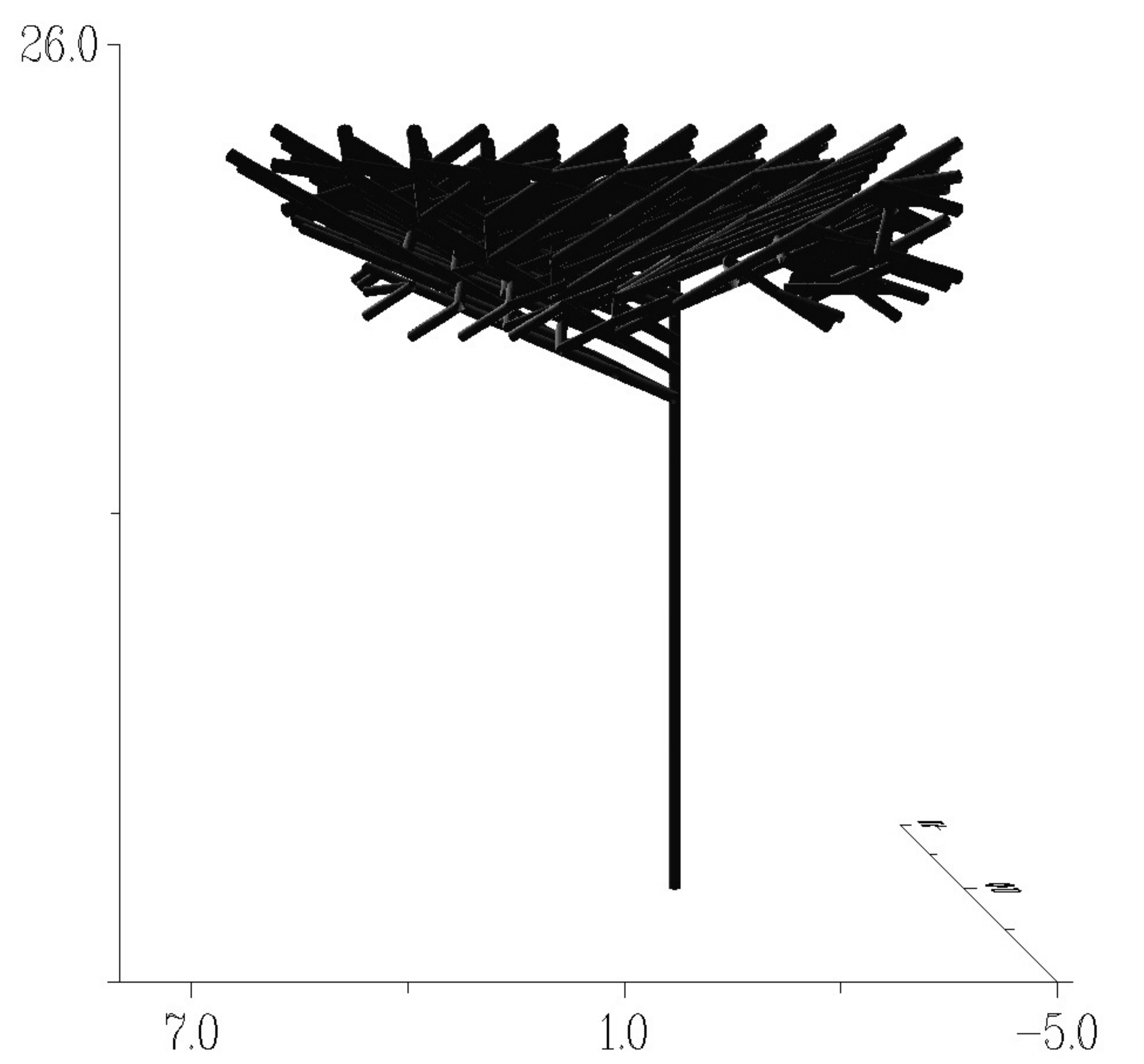}\caption{\label{fig:Fig6}Using the same parameters as Fig.~\ref{fig:Fig5}
with $a=2$. The direction of the children are not chosen randomly.
The tree on the right uses the apical dominance scheme described in
Sect.~\ref{subsec:Non-spatial-tree} at 10\%.}
\end{figure}
 We observed that the addition a spacial component and some spatial
exclusion is enough to simulate a self-pruning tree for some simple
resource redistribution scheme (Fig.~\ref{fig:Fig5}) in a way similar
to the apical scheme did Fig.~\ref{fig:3}. Though, with the directions
of growth chosen randomly, the tree would not grow straight.

Instead of deciding randomly the direction of children, the parents
could make their decision according to a precise algorithm. The leftmost
tree of Fig.~\ref{fig:Fig6} shows such simulation: each parent creates
its first child in the same direction it is pointing toward, then
the other children are birthed in a balanced way: it tries to avoid
having all the children facing the same direction. The ``root''
branch is considered as being vertical, so its first child will be
in the vertical direction. We must remind that, when several extremities
have the resource to birth children, they do not decide the locations
of their future children simultaneously, instead an ordered list is
followed, as such branches that are high on the list get to book their
``favored'' locations before the other ones. We also need to remind
the list is ``rigid'': for example, given 2 siblings A and B with
a parent named C, if A is the ``first'' child of C then the children
and all the descendants of A will be higher on the list than B or
any of its descendants, this is true throughout the simulation. Thus,
we could have expected the tree to grow straight since the vertical
branches from the ``root'' are ``first'' children of their respective
parents. To explain the fact the tree systematically grows laterally
(tree on the left in Fig.~\ref{fig:Fig6}), we could believe it is
due to a lack of space in the center. We may speculate that if our
3D-space did not have this cubic metric, the tree could have grown
straight but the cubic metric advantages diagonal directions.

Regardless of this assumption, to elicit a straight growth or more
generally speaking to control the form of the tree in a desired way,
a more direct approach seems necessary as seen in the second tree
of Fig.~\ref{fig:Fig6} where we use the ``apical dominance'' approach
presented in Sect.~\ref{subsec:Non-spatial-tree} on top of the ``child's
location decision'' algorithm. Even then, this only managed to control
the trunk, the crown and the topology of the tree is still similar
to the leftmost tree of Fig.~\ref{fig:Fig5} where all the leaves
are clustered, and unlike the tree on the right of Fig.~\ref{fig:Fig5}
with its multiple clusters. So we want to seek other approaches to
control the tree growth: forcing a straight growth, controlling the
shape etc. For the following exploration models, we start from the
``randomly-generated direction'' model again (see Sect.~\ref{subsec:Children-generated-in})
and append to it factors that could control the growth.

\subsection{Gravitational loads\label{subsec:Gravitational-loads}}

Following the call in Sect.~\ref{subsec:Non-random-directions} to
investigate how to control this dynamical system, we seek to add some
parameters or factors to the model. Let us seek some intuitive factors.
Since the model is inspired by biological tree a natural factor that
could control its shape would be mechanical constraints. In the literature,
many models of mechanical constraints on branches and trees are based
on simple beam theory arguments\cite{Cannell1989,Niklas1999a,Spatz2000a,Eloy2011a}.
Though wind loads are important\cite{Spatz2000a,Eloy2011a}, for
a simpler factor we could limit ourselves to gravitational loads only:
the base of a branch will be subjected to a stress induced by the
weight of all the branches and leaves it is supporting. When a beam
is bent some parts of it is compressed whereas other parts is stretched,
and in-between is a neutral axis experiencing no stress. The further
from the axis an area is the more stress it experiences such that
the surface of the branch has the most stress. This maximum bending
stress occurring at the surface is the quantity of interest. Let $\sigma_{s}$
this surface stress, in the simple model we will use\cite{Cannell1989,Niklas1999a},
we associate a single constant $\sigma_{b}$ to the wood the tree
is made of. This parameter represents the stress limit a branch can
take: the branch breaks if the stress $\sigma_{s}$ is higher than
$\sigma_{b}$. Furthermore we have the formula:
\begin{equation}
\sigma_{s}\propto M/d^{3}\label{eq:sigmas}
\end{equation}
 where $M$ is the bending moment and $d$ the diameter of the branch\cite{Cannell1989,Niklas1999a}.

The implementation of this theory into our model is as follow: during
``flux down'', and after a branch grows in size to match the number
of extremities it supports, the moment of force resulting from the
combined weight of all of its descendants is calculated. (We simply
define the mass of a branch as ``equal'' to its volume.) From there,
given a constant parameter $S_{b}$, which is our equivalent of $\sigma_{b}$,
the branch breaks when 
\begin{equation}
S_{b}<T/V^{\gamma}\label{eq:2.5}
\end{equation}
 where $T$ is the moment of force, $V$ the volume of the branch
and $\gamma$ will be put at $3/2$. Indeed, contextualizing our model
within the plant-analogy, what we call ``branches'' are segments
of actual branch of some fixed length. With the ``length'' being
held constant, $V$, the volume of the branch, only scales to its
cross-section area which itself scales like the diameter squared resulting
in $d^{3}\propto V^{1.5}$. Thus, $\gamma=3/2$ allows us to mimic
Eq.~\ref{eq:sigmas}. Another important remark regarding the model
is that the branches closest to the extremities will break in ``priority'':
if both branches $A$ and $B$ have reached the breakage limit $S_{b}$
and $B$ is a descendant of $A$ then $B$ will break first, afterward
the moment of force for $A$ would be recalculated accounting for
the disappearance of $B$. $A$ would not break if the new calculation
puts it below the limit $S_{b}$. Other than that: the direction for
the children are still chosen randomly, and we will simply hope that
the mechanical constraint when combined with some reward redistribution
scheme would allow the tree to take shape on its own.

Unfortunately the simulations end up with a growth similar to what
was described in Sect.~\ref{subsec:Children-generated-in}, self-pruning
and trunk appear but they do not grow vertically and instead can take
some random directions. We could have hoped for the gravitational
load factor introduced in this Section to straighten up the tree by
progressively removing lateral branches and ``sculpting'' a vertical
tree. However, this is not the result we get. The gravitational factor
does not largely impact the form of the crown in many cases, with
he exception of the ``maintenance'' redistribution case with very
high value of $b$. We obtain the same pattern as described in Sect.~\ref{subsec:Children-generated-in}
with respect to the reward redistribution scheme and maintenance scheme
at low $b$. The difference is that the tree collapses earlier. For
example, using the ``reward'' scheme and the values of $a$ that
induce self-pruning, we would get a tree growing similarly to the
leftmost tree in Fig.~\ref{fig:Fig5} which leans too far in one
direction, then the mechanical factor would directly break the branches
at the base of the crown leaving just a shorter and naked trunk. A
quirk of the model is that a collapsed trunk would typically revives
because all extremity produces $p_{0}$ resource and therefore even
a tree reduced to a single trunk would immediately being able to produce
$p_{0}$ resource and regrow a crown which will end up brutally collapsing
for similar reason and restart this cycle a number of times. On the
other hand, as we implied, the ``maintenance'' scheme at high $b$
that, in Sect.~\ref{subsec:Children-generated-in}, gave us the multi-cluster
tree in Fig.~\ref{fig:Fig5} is affected, for the worse: the mechanical
failures seem to prevent the self-pruning into the complex multi-cluster
shape. The tree stays in a bush-like state.

A more thorough discussion on the problem is done in Sect.~\ref{subsec:Explaining-sudden-collapses}.

\subsection{Light interception factor for productivity\label{subsec:Light-interception-factor}}

\begin{figure}
\includegraphics[width=100bp]{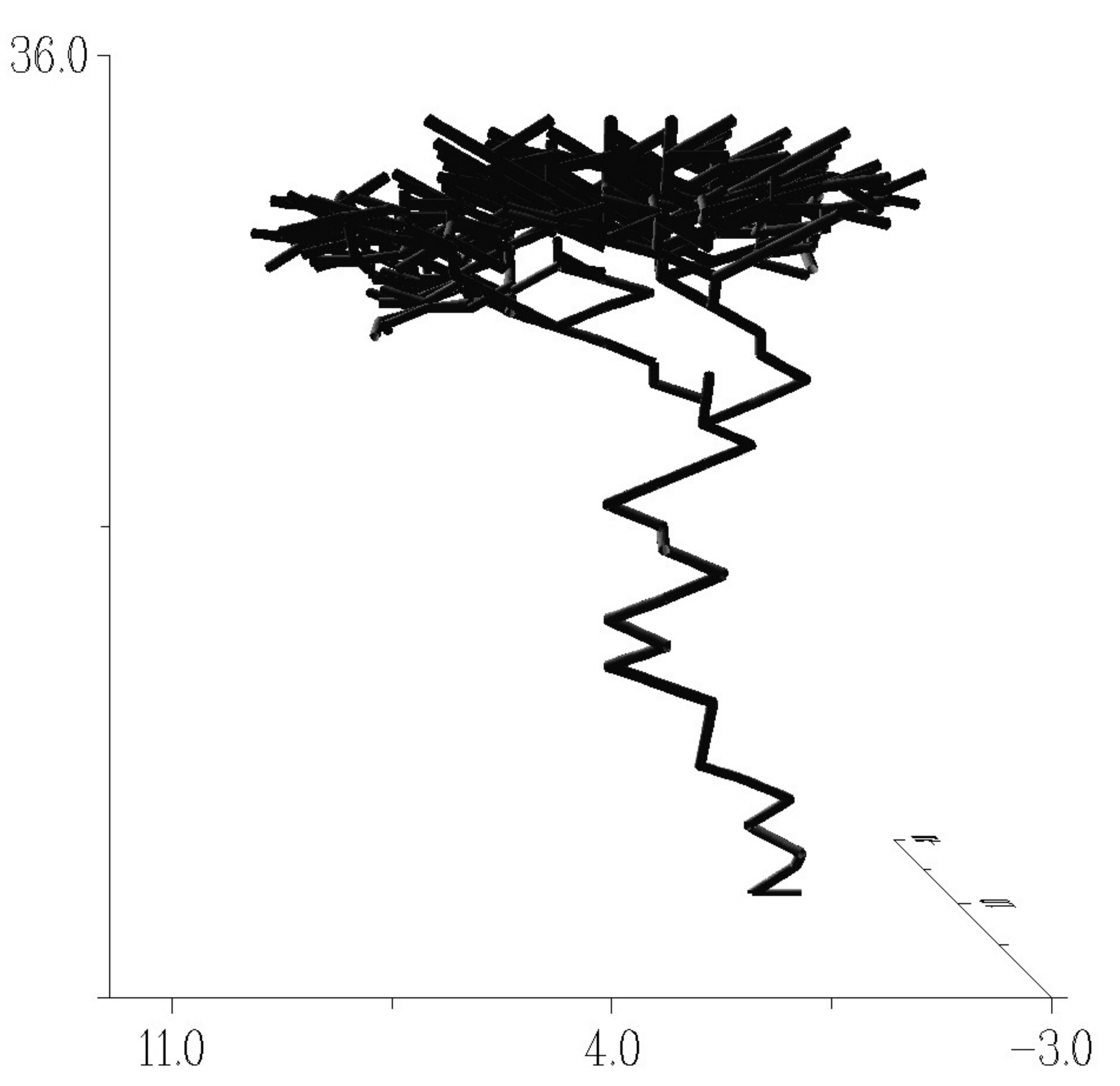}\includegraphics[width=100bp]{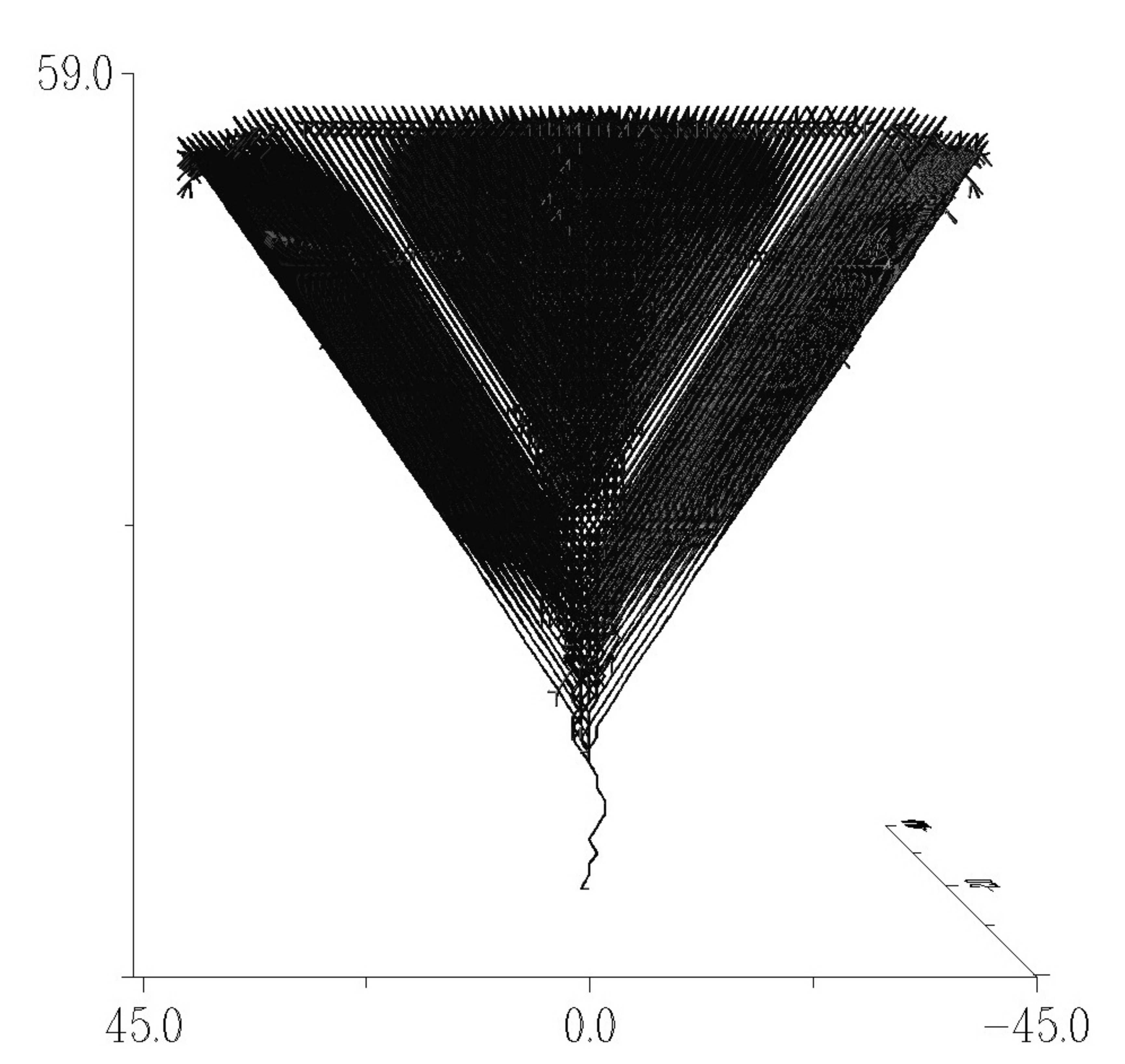}

\caption{\label{fig:Fig7} Using the parameters in Fig.~\ref{fig:Fig5} with
some exceptions: $p_{0}=52$, The tree on the left uses $a=2$ while
the right one uses $a=1$. These tree are simulate with the inclusion
of a light interception factor whereby the leaves/extremities shaded
by other branches from the ``Sun'' can not create resource.}
\end{figure}
 Starting again from Sect.~\ref{subsec:Children-generated-in} again,
and searching for a different factor from the mechanical one in Sect.~\ref{subsec:Gravitational-loads}
that could drive the shape of the tree, we choose to implement some
light interception scheme. Unobstructed extremities would still produce
$p_{0}$ resource while the ones hidden behind other branches would
produce less due to less exposure to the ``Sun''. In real life situations
the interception of light by leaves is complex as the rays are not
simply in the direction of the Sun: light diffusion through the atmosphere
called sky radiation and light scattering by clouds or other leaves
are all non-negligible sources of radiation\cite{Erbs1982,VERHOEF1984,Bosquet2016}.
The direction of light is particularly important as models to explain
the shape of real life trees based on that aspect has been proposed
in the past\cite{Niklas1994a,Duchemin2018}. Despite the complexity
of the issue, the present model will only account for direct sunlight
and ignore diffusion and scattering. Sunrays will hit the X-Y plane
with an angle $\theta$ from $0$ radian to $\pi$ representing the
diurnal cycle. When a ray hit a position occupied by a branch, it
is stopped without scattering or reflexion. More concretely, at the
start of each generation, an uniform sampling between 0 rad and $\pi$
rad is performed for each extremity. An extremity that received all
the rays from the angles sampled would produce $p_{0}$ resource.
Otherwise it would receive a fraction of $p_{0}$ equal to the fraction
of rays it received. In other words, the production is proportional
to the amount of unobstructed angles. Fig.~\ref{fig:Fig7} is obtained:
children's locations are chosen at random like in Sect.~\ref{subsec:Children-generated-in}
and the gravitational aspect introduced in Sect.~\ref{subsec:Gravitational-loads}
is turned off. We consistently obtain a straight trunk despite the
random nature branch generation. And unlike the gravitational loads
model the crown is not regularly destroyed, instead the pruning is
progressive and constantly shape the tree. The foliage seen in the
leftmost tree of Fig.~\ref{fig:Fig7} is much smaller than what we
saw in the leftmost tree of Fig.~\ref{fig:Fig5} but the behavior
is the same: after the initial self-pruning and trunk formation phase,
the length of the trunk will grow but the number of leaves is somehow
maintained close to constant and we do not have the multi-cluster
leaves of the rightmost tree in Fig.~\ref{fig:Fig5} but the usual
mono-cluster. Slight variation in the model such as weighting each
angle $\theta$ differently, instead of each sunray angle being counted
the same, (for example a weight of $\sin(\theta)$) does not have
a noticeable effect on the shape or behavior of the tree. If, instead
of sampling $\theta$ from $0$ to $\pi$, we only include vertical
rays, we get the second tree of Fig.~\ref{fig:Fig7} instead of a
bush-like tree for the ``reward'' scheme at low $a$ values.

Despite being able to force a vertical growth, the shape of the tree
or its behavior do not show a large amount of diversity as Fig.~\ref{fig:Fig7}
shows all the new shapes and behaviors. On the other hand the multi-cluster
leaves shape obtained previously (Fig.~\ref{fig:Fig5}) with the
``maintenance'' scheme at high value of $b$ has disappeared. This
particular growth seems quite vulnerable as it also broke in face
of the ``gravitational load'' factor.

\section{Numerical and theoretical survey \label{sec:A-static-description}}

In this Section, we perform some analytical calculations and, try
to analyze and summarize the results of the simulations as well as
main takeaway from the model.

\subsection{Tree growth and spatial embedding\label{subsec:Tree-growth-and}}

We analyze and summarize the results of the simulation.

\subsubsection{Description of tree growth}

\begin{figure}
\includegraphics[width=240bp]{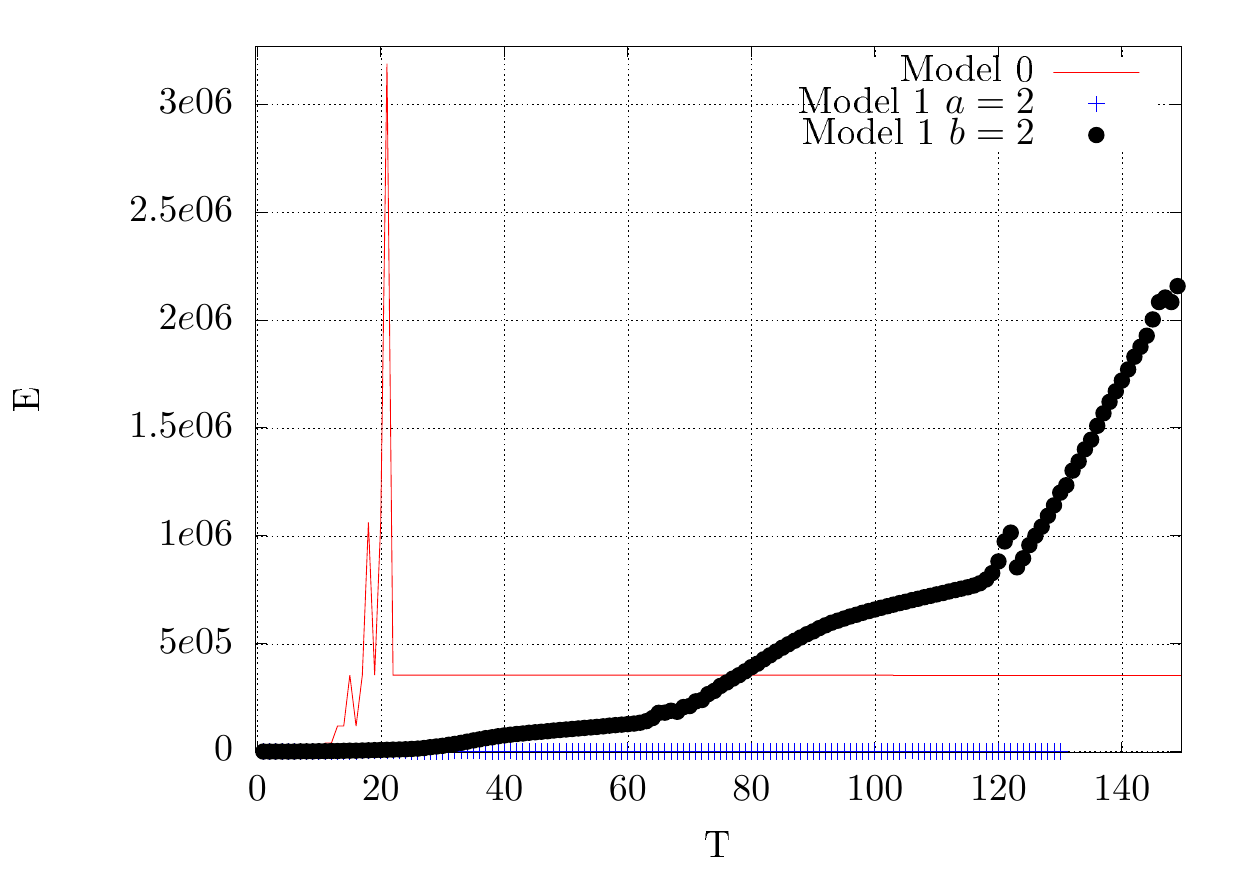}

\includegraphics[width=240bp]{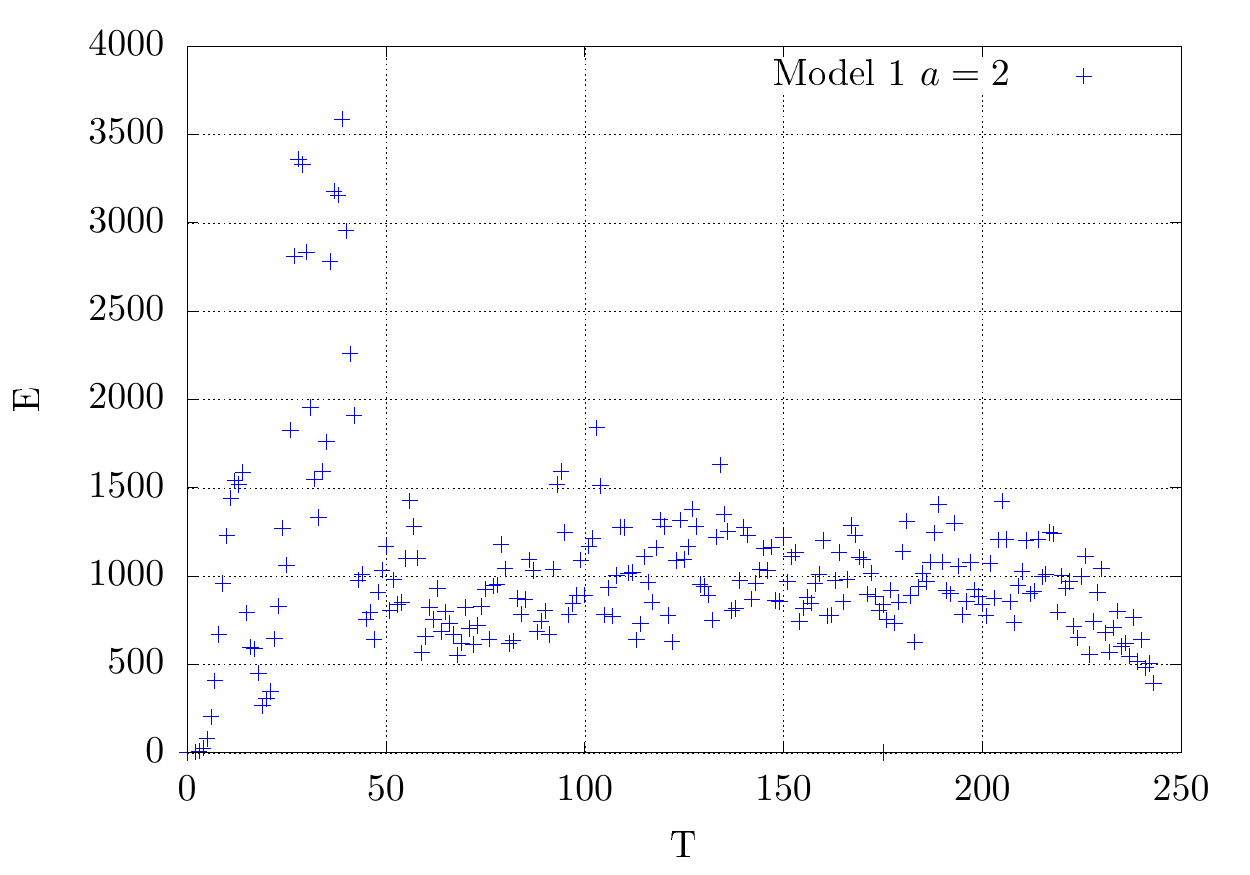}

\caption{\label{fig:Fig8}The first Fig. uses $p_{0}=9.6$, the second one
uses $p_{0}=12$ and both are under $m_{0}=1$, $C_{m}=1$, $N_{max}=3$.
The first one showcases the evolution the number of extremities $E$
with time $T$ in three trees: what we call Model 0 is a symmetric
tree (no ``reward'' or ``maintenance'' scheme and no ``apical
dominance'' but instead each parent shares equally its flux) and
Model 1 is the spatial embedding with a random selection for the children's
direction. The second Fig. is a zoom for the Model 1 and using the
reward redistribution scheme at $a=2$.}
\end{figure}
 Sect.~\ref{subsec:Children-generated-in} showed the asymmetry created
by simply embedding in space the tree was enough to trigger the kind
of self-pruning which happens when we force resource to be distributed
to a some special branches (``apical dominance'' in Sect.~\ref{subsec:Non-spatial-tree}).
Beyond looking at the formation or lack of trunks, we can also look
at the number of extremities each tree has. Fig.~\ref{fig:Fig8}
shows some differences between three trees: the one noted as Model
0 is a tree that is not embedded in space and where each parent share
the resource equally among their children, and Model 1 are trees in
space using either the ``reward'' distribution, whereby the resource
are distributed to the more productive children, or the ``maintenance''
distribution which focuses on each child's needs. The evolution of
these three trees are seen in Fig.~\ref{fig:Fig8}. Both the ``symmetric''
tree (Model 0) and the ``reward'' tree (zoomed in the second plot
of Fig.~\ref{fig:Fig8}) end up in a stable phase after an initial
collapse that happens quite fast. For the ``reward'' tree the collapse
corresponds to the pruning and trunk formation events. On the other
hand, the ``symmetric'' tree does not form trunk, the collapse event
is just a brutal collapse into an equilibrium state. Something that
is not shown in the second plot of Fig.~\ref{fig:Fig8} is that despite
a numbers of extremities oscillating, the tree itself does not: the
trunk continues to grow in length at a constant rate while the numbers
of leaves oscillates. This contrasts with the symmetric tree which
is actually in an equilibrium state. The ``maintenance'' tree is
the only one that grows its number of leaves for a long time, it usually
brutally collapses and loses most if not all its branches. If a few
branches remain it sometimes restarts the same growth as the one it
had at the start of the simulation, but in either cases equilibrium
or stationary states are not reached.

\subsubsection{Interactions between space and growth}

The system is not very sensitive to randomness. Indeed, the randomness
generated from choosing the children's directions does not affect
the behavior for the ``reward'' and ``maintenance'' tree as measured
by values like number of extremities or length of the ``trunk''.
The main difference between simulations is the direction or path the
trunk takes, however, the direction set aside, the overall shape is
not modified by repeating the same simulations with different randomly
generated directions. The only possible caveat to this last statement
is the ``maintenance'' scheme using large $b$ that provided us
with the multiple cluster of leaves we obtained in Fig.~\ref{fig:Fig5}.
The system is not very sensitive to change in the parameters as we
get the same patterns by changing the parameters slightly. And for
parameters unrelated to redistribution schemes, even large changes
usually have weak effects: the behaviors are identical but there may
be slight changes in numbers of leaves as seen in Fig.~\ref{fig:Fig9}.

This stability is also observable from the simulations done in Sect.~\ref{subsec:Non-spatial-tree}
in which children's directions where not random anymore: the trunk
curved less but the topology and overall behaviors were similar. Then
when we added ``apical dominance'' to the ``reward'' tree with
non random directions, the only striking topological difference was
a lower average number of extremities for the crown (Fig.~\ref{fig:Fig6}).
Increasing the value of $a$, the exponent determining how much we
focus on reward (cf. Eq.~\ref{eq:3}), has initially a large impact
but its effect vanishes quickly (Fig.~\ref{fig:Fig9}).

In a way we could conclude there is little interactions between space
and growth since how the tree growth explores the space does not noticeably
retroact on said growth. We could further test that statement: let
us call ``ground'' the plane below the initial branch, the root
of the tree. We can remove the ``ground'' and see the effect of
it: it had also no noteworthy effect on the growth.

Of course it does not mean the 3D embedding did nothing: it limits
the numbers of branches a tree may have at any given time which is
very apparent when comparing the number of branches or extremities
(Fig.~\ref{fig:Fig8}) between the two models. And the 3D embedding
allowed the sustained symmetry breakage needed for the redistribution
scheme to form an asymmetric tree. Though, the non-impact of the ``ground''
makes it clear the asymmetry is not caused by the extremities near
the ground being disadvantaged.

\subsubsection{Influence of the parameters}

\begin{figure}
\includegraphics[width=220bp]{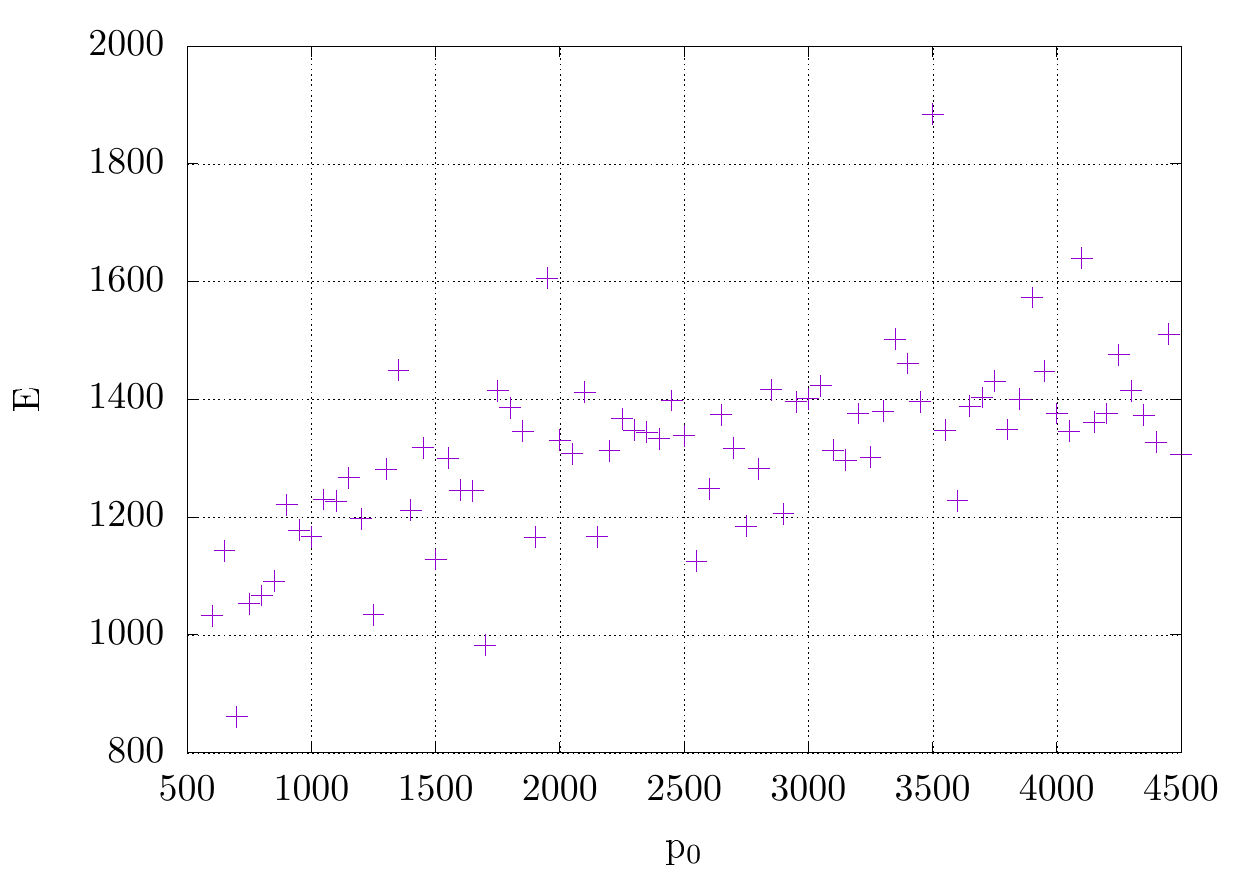}

\includegraphics[width=220bp]{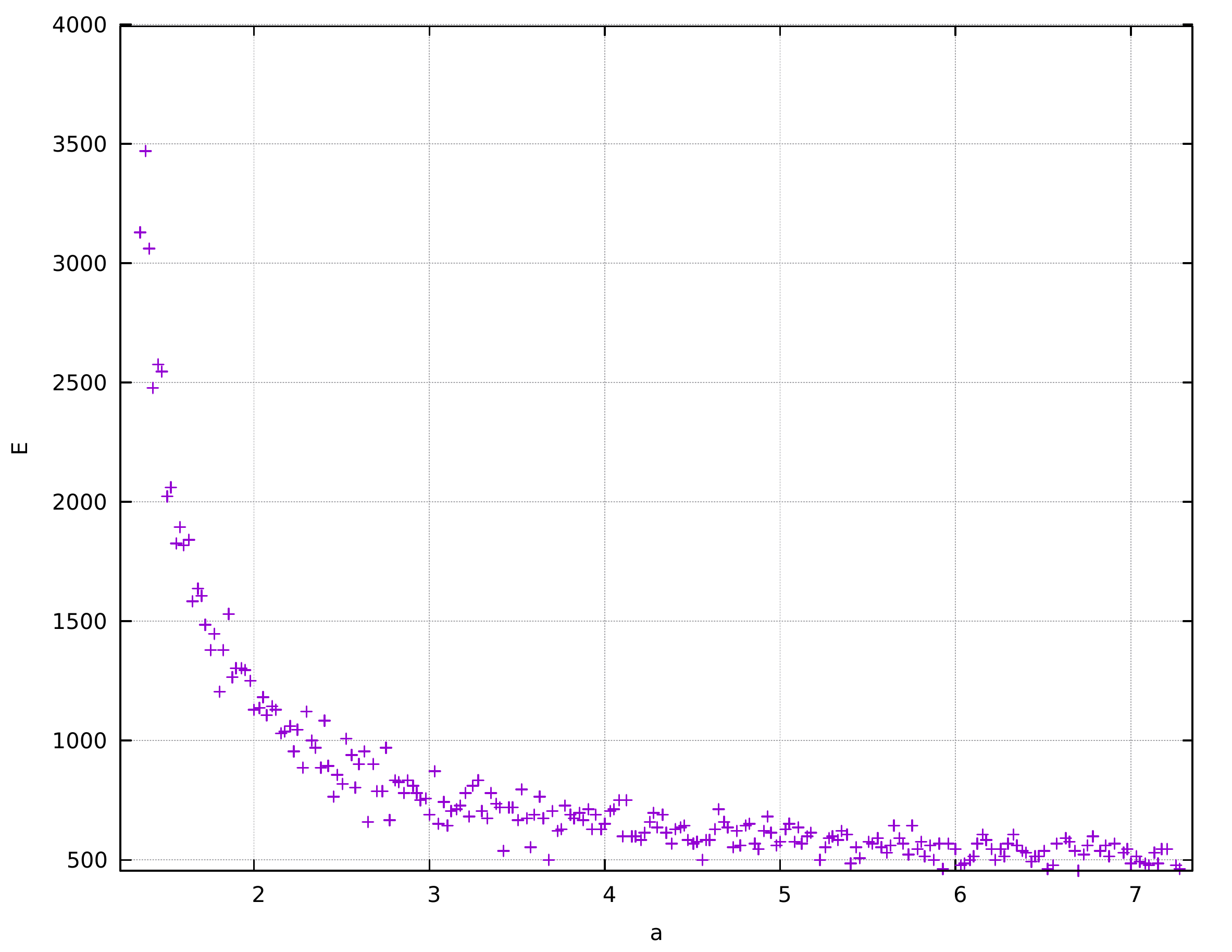}\caption{\label{fig:Fig9} Parameters not mentioned on the plots holds the
same values as previous plots. $E$ is the average number of extremities
a simulated tree had after self-pruning. Indeed, when using the ``reward''
scheme and a sufficiently high value of $a$ then, after some transitory
phase, the tree will self-prune and from there the number of extremities
will only oscillate around a number and $E$ is that number. We plot
$E$ against different values of $p_{0}$ then different values of
$a$.}
\end{figure}
 As we showed: when using the ``reward'' resource distribution scheme,
after some self-pruning, the number of leaves stays stable (Fig.~\ref{fig:Fig8}).
This stable number of extremities is used to characterize each tree
grown in a given set of parameters (Fig.~\ref{fig:Fig9}). We also
mentioned that how the tree growth explore the space or whether there
is a ``ground'' or a wall are not an important factor. The redistribution
scheme is the main factor that drives how the tree grow. We can see
its effect in Fig.~\ref{fig:Fig9}. However in both cases, the system
reacts continuously with respect to the parameters.

There is however one caveat against this narrative of simplicity and
stability presented so far. The caveat are trees appearing when using
the ``maintenance'' scheme and large values of $b$ (like Fig.~\ref{fig:Fig5}).
When measured through the lens of ``number of extremities'' or ``length
of the trunk'', we get trees that are no less stable than the ``reward''
trees when repeating simulations or changing parameters, however it
is possible a more detailed investigations of the topology could show
differences between different simulations: for example, by trying
to count the number of clusters of leaves these trees have. The third
tree in Fig.~\ref{fig:Fig5} certainly shows a network complex enough
it should not be described solely by its number of extremities or
trunk. Furthermore, unlike the ``normal'' mono-cluster trees, these
multi-cluster trees disappear completely when we applied gravitational
loads or light interception (Sect.~\ref{subsec:Gravitational-loads}
and \ref{subsec:Light-interception-factor}) setting them as structures
easily affected by perturbations.

\subsection{Effects of the redistribution schemes on trunk formation\label{subsec:Influence-of-the}}

\begin{figure}
\includegraphics[width=240bp]{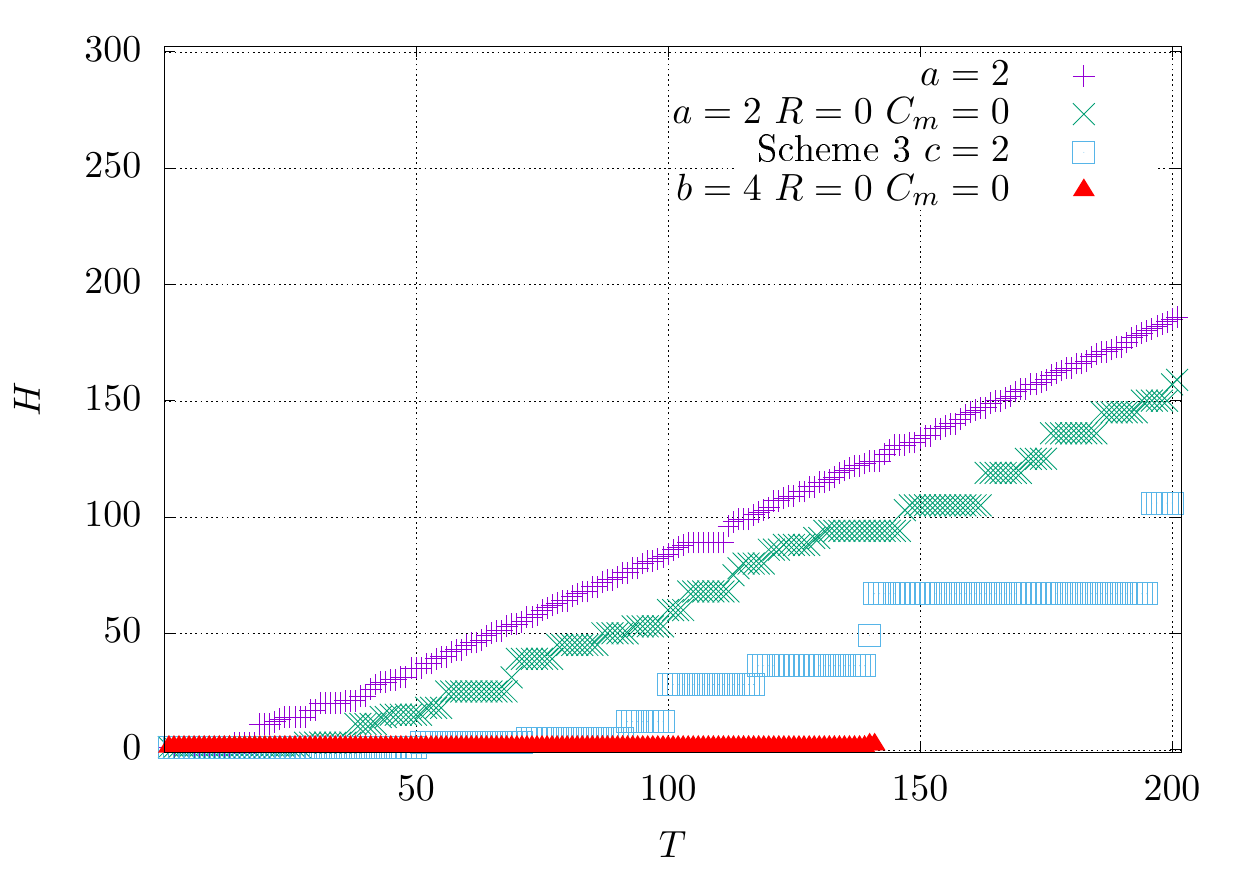}

\caption{\label{fig:Fig10} How the length of the trunk of a tree, noted $H$,
grows as time $T$ passes. We first plot a normal ``reward'' tree
at $a=2$. Then in order to showcase the effect of ``leftovers''
and ``non-decay of volumes'' (see Sect.~\ref{subsec:Influence-of-the})
on self-pruning, we simulate trees lacking these factors. The second
plot has no leftover (noted $R=0$). For the third plot we use a custom
redistribution scheme similar to the ``reward'' scheme but, instead
of computing the total flux each child gave during ``flux down'',
parents compute the flux they would have received if $R=0$ and $C_{m}=0$
and used this hypothetical flux to decide on the share for each child.
The last plot uses the ``maintenance'' scheme (i.e. the ``non-decay
of volumes'' is turned on).}
\end{figure}

In Sect.~\ref{subsec:Non-spatial-tree} we alluded to the idea that
when $\alpha=1/2$ we could expect a tree using the ``reward'' distribution
with $a=x$ would create the same tree as one using the ``maintenance''
distribution with $b=2x$. Indeed, the ``reward'' scheme favors
the more productive branches while the ``maintenance'' one advantages
the bigger branches, since the volume of a branch is, under normal
circumstances, ``equal'' to the number of extremities originating
from it, then ``maintenance'' just be a rescaling of ``reward''
when we ignore some complications. This rescaling is however not observed
(Fig.~\ref{fig:Fig5}): the behavior between ``reward and ``maintenance''
is so different it can not be reduced to a $b=ya$ relation let alone
$b=2a$. Indeed, despite formation of trunk for $a=2$, no trunk appears
for $b=4$ (Fig.~\ref{fig:Fig10}), and while some trunk appears
for $b=6$ and beyond their length is not comparable to their counterpart
and the topology is much more complex. The evolution of the number
of extremities does not stabilize for $b>6$, unlike the ``reward''
trees: the plot of the number of extremities with respect to time
is similar to the $b=2$ case drawn in Fig.~\ref{fig:Fig8} (the
difference is over the number involved: higher $b$ shows lower numbers
but the shape of the plot is similar). Thus, on the one hand, no notable
``interaction'' between growth and space has been noted and the
system is resilient to randomness (Sect.~\ref{subsec:Tree-growth-and}),
on the other hand, some differences between the ``$a=x$'' and ``$b=2x$''
redistribution schemes end up creating totally different trees.

We must list the factors that explain the difference between the ``reward''
and ``maintenance'' schemes. First are ``leftovers'' from previous
cycles. Indeed, considering a branch can only create $N_{max}$ children,
if it possess an amount higher than $N_{max}\times C_{r}$ there will
be some leftover resource that will be carry in the next ``flux down''
and will make branches looks more productive to its parent. But another
source of leftover, and one that would asymmetrically affect different
branches, is spatial limitation: a branch unable to create as much
children as it could have because the adjacent locations are occupied
will not be able to use up its resource which in turn will be carried
over as leftovers.

Aside from leftovers, another factor is the cost of growing volume
which amounts to $C_{m}$ multiplied by the amount of volume a branch
creates. We remind that the volume of a branch needs to be equal to
the number of extremities it supports (Leonardo's rule mentioned in
Section.~\ref{subsec:Model}). So let us assume at generation $T$,
a branch supports $E$ extremities, its volume is also $E$, but at
$T+1$, it grows to $E+A$ extremities. From the ``reward'' redistribution
formula Eq.~\ref{eq:3}, and assuming no ``leftovers'', the share
the branch will get from its parent will be:
\begin{equation}
\begin{array}{ccc}
F_{T+1} & = & \dfrac{\left(p_{0}(E+A)-AhC_{m}\right)^{a}}{Z_{T+1}}\\
 & = & \dfrac{\left(E+A(1-h\frac{C_{m}}{p_{0}})\right)^{a}}{Z_{T+1}/p_{0}^{a}}
\end{array}\label{eq:5}
\end{equation}

The $AhC_{m}$ term assumes all the $A$ extremities added at generation
$T+1$ are at a distance $h-1$ of the branch (distance in the graph-sense).
But if we had $A_{1}$ added at distance $h_{1}-1$ and $A_{2}$ added
at distance $h_{2}-1$ the term is to be replaced by $(A_{1}h_{1}+A_{2}h_{2})C_{m}$,
and this is generalizable to sums with more than 2. We can convince
ourselves of the $AhC_{m}$ term by imagining the $A=1$ case: one
branch was added at a distance $h-1$, this results to the parent
of this branch needing to get 1 more unit of ``Volume'', the grandparent
would also need 1 more Volume, etc. until reaching the initial branch.
As such the total Volume increased is $h$. By induction, we end up
with $AhC_{m}$. Now, looking at Eq.~\ref{eq:5}, we can easily see
it should have mitigating effect on increased inequalities between
children. Indeed, if we compare a child that gained $A$ with one
that did not, had the term $AhC_{m}$ not been present (which is the
case in the ``maintenance'' scheme) then the former would gain even
more share from the parent, moreover the share can be even lower if
$h>C_{m}/p_{0}$ in Eq.~\ref{eq:5}.

Finally a third factor making the ``maintenance'' and ``reward''
schemes different from one another even when ``$b=2a$'' is the
fact volume can not decay. A branch that supported $E$ extremities
at generation $T$ but only $E-A$ extremities at $T+1$ will still
have a volume of $E$ despite only ``producing'' $(E-A)p_{0}$ at
best. This effect should mitigate inequalities in the ``maintenance''
scheme as a loss of leaves do not result in a lesser share; but it
will be a source of inequalities for the ``reward'' scheme.

How the three factors compete with one another to create the result
we see can be summarize by Fig.~\ref{fig:Fig10} which depicts how
the length of the trunk evolves with time. We start from a ``normal''
tree with $a=2$ then we simulate the case $C_{m}=0$ which is not
shown in Fig.~\ref{fig:Fig10} as the plot is identical to the ``normal''
tree, indicating the effect of $C_{m}$ is weak compared with the
other factors. Then we forbid ``leftovers'' by forcing the reserve
$R$ to be always $0$. The Figure allows us to see how we go from
the ``reward'' trees with their growing trunks to the ``maintenance''
ones which do not self-prune.

The takeaway is $C_{m}$ is not important for trunk formation while
``leftovers'' have some effect. But the ``non-decay'' is the main
factor for self-pruning or its absence when comparing $a=2$ with
$b=4$. By definition the ``non-decay of volume'' factor only differentiates
the two schemes after branches start to fall: it means the self-pruning
in ``reward'' trees is not caused by groups of branches progressively
getting more leaves and monopolizing more shares instead it is brutal
incident such as death of probably ``basal'' branches that starts
and drives it. On the other hand, the pruning that happens in ``maintenance''
scheme with the large values $b>6$ (Fig.~\ref{fig:Fig5}) is probably
driven by such progressive monopolizing process. This could explain
the differences in topology between those two self-pruned trees, though
it is not very clear why the ``progressive monopolizing'' creates
multiple clusters while the other process favors mono-clusters.

\subsection{Gravitational load\label{subsec:Explaining-sudden-collapses}}

As explained in Sect.~\ref{subsec:Gravitational-loads}, we fail
to control the shape of the tree in any meaningful way by adding a
gravitational load factor to the trees generating children in random
directions: we do not obtain the vertical trunk we expected. Rather
than driving the pruning and straightening the tree, breakage from
gravitational loads are mostly punctual and cause a brutal collapse
of the tree only once it went too far in a direction. The exception
is when we use extremely low threshold of breakage to the point that
a branch going two or three units of space on lateral directions would
break: in such circumstances the tree would simply be a straight trunk
with two or three branches that are extremities at the very top, and this is
not the kind of tree or growth we are looking for either. Attempts
to give weight to leaves (i.e. extremities weighting extra units)
does not yield results different than the ones shown so far.

\begin{figure}
\includegraphics[width=140bp]{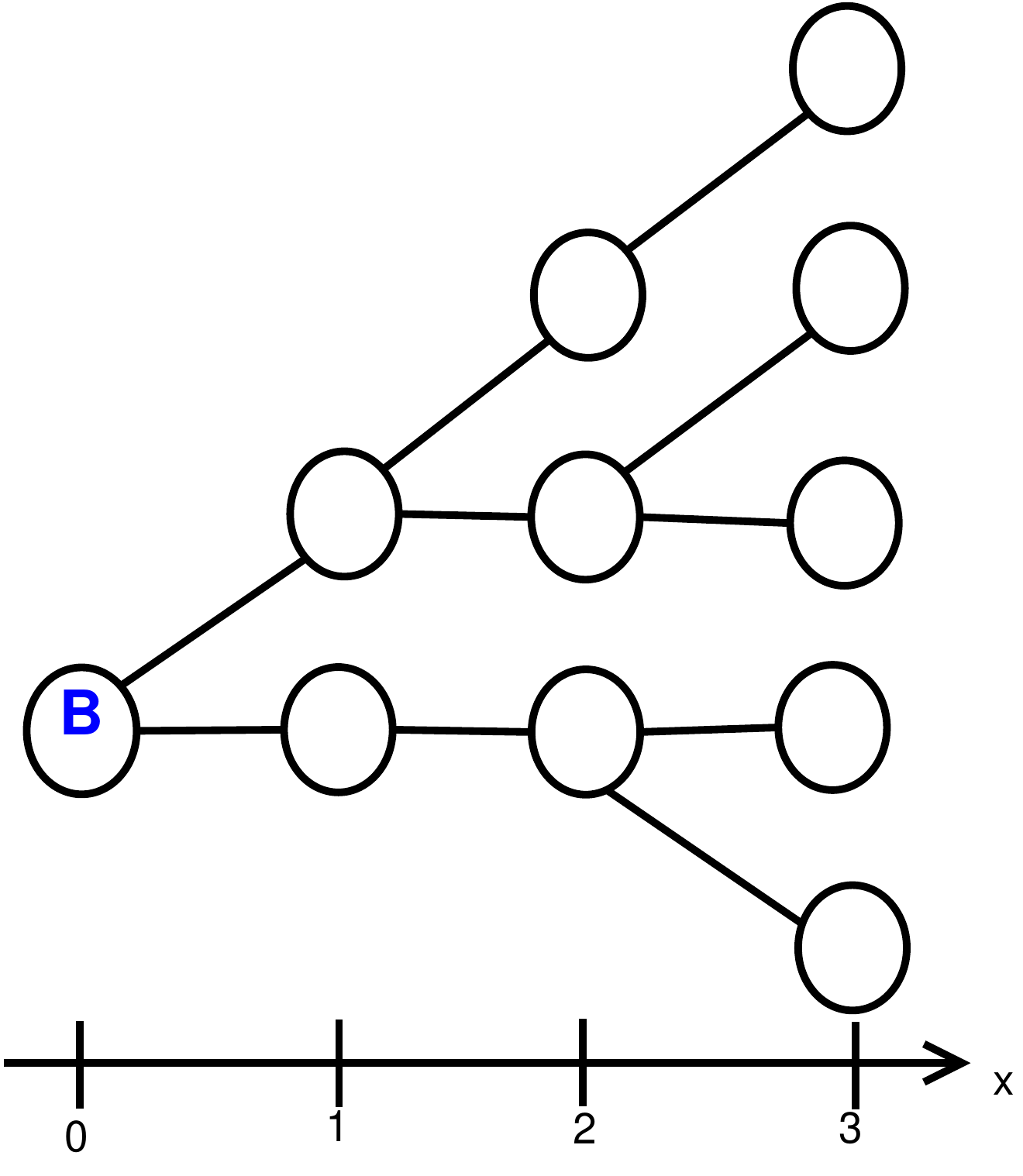}

\caption{\label{fig:Fig11}A sketch of a part of a tree. Starting from some
branch $B$, all children are located on the right of their parent.
This simple example serves as a starting point to determine how branches
break under gravitational loads (see Sect.~\ref{subsec:Explaining-sudden-collapses}).}
\end{figure}
Let us try to understand how branch breakage works by analyzing simple
case such as in Fig.~\ref{fig:Fig11}. For simplicity the example
is in 2D. For the calculations to be simple, branch $B$ has all its
children on its right, the grandchildren are on the right of the children
etc. And we wish to determine whether the branches tend to break near
the extremities or near the base.

We remind Eq.~\ref{eq:2.5}: $S_{b}<T/V^{\gamma}$ and the bending
moment $T$ is the combined volume of all the descendants of branch
$B$ multiplied by the distance between the barycenter of the descendants
and the branch $B$ when both these points are projected onto the
X-Y plane. Another assumption for the Fig.~\ref{fig:Fig11} example
is: all the extremities descending from branch $B$ are at the same
``distance'' $d$ (distance in a graph-theory sense) from it. Leonardo's
rule needs to be true (volume of a branch equals the number of extremities
it supports), meaning the total volume of the descendants of $B$
is $V_{B}d$ where $V_{B}$ is the volume of $B$. This can be easily
verified in simple example like in Fig.~\ref{fig:Fig11} and the
arguments for it are similar to the derivation of the $AhC_{m}$ terms
in Eq.~\ref{eq:5} but with $d$ taking the role of $h$. If the
extremal descendants are not all at a distance $d$ but some are at
a distance $d_{1}$ while others are at $d_{2}$ the formula is changed
like we did with Eq.~\ref{eq:5}. Noting $\Delta_{B}$ the distance
(in the euclidean sense) between branch $B$ and the barycenter when
both are projected onto X-Y plane representing the ground, we have
the stress $S=T/V_{B}^{\gamma}$ equal to: $S=\dfrac{d\Delta_{B}}{V_{B}^{\gamma-1}}$.
Assuming each child is strictly to the right of its parent (Fig.~\ref{fig:Fig11})
$\Delta_{B}=(d+1)/2$: indeed, if $x_{B}$ is the $x$ position of
$B$, then all its children are at $x_{B}+1$, but the sum of volumes
of all children of $B$ is $V_{B}$ because of Leonardo's rules. Applying
the same reasoning to the grandchildren then grand-grandchildren etc.
leads to the barycenter being at $x_{B}+(d+1)/2$. The formula for
$\Delta_{B}$ relies on the extremities being located at the position
$x=x_{B}+d$. But even assuming a general setting, $\Delta_{B}\propto d$
could be close to reality as long as the descendants approximatively
grow toward a lateral direction. The end result would be: 
\begin{equation}
S\propto\dfrac{d^{2}}{V_{B}^{\gamma-1}}\label{eq:6}
\end{equation}
 The last term needing to be expressed as a function of $d$ is $V_{B}$.
Once such a relation is given we obtain a function $S(d)$. If the
children of the branch $B$ also follows the assumption we needed
to established the $S(d)$ formula (i.e. starting from branch $B$
there must be a somewhat self-similar growth) then the stress $S$
on a child of $B$ will be close to $S(d-1)$, this means $S(d)$
would determine how gravitational loads break branches. There is not
an unique way for $V_{B}$ to vary as a function of $d$ (reminder:
$V_{B}$ is also the number of extremities from $B$). Assuming a
form of self-similarity in the growth from branch $B$ and that it
grows laterally, we could approximate $V_{B}\propto d^{\beta}$ with
$\beta$ an undetermined exponent. In a configuration like Fig.~\ref{fig:Fig11}
where we have a ``lateral triangular'' growth $\beta\simeq1$: the
angle at branch $B$ is constant, therefore if $d$ grows it will
linearly affect the length of the base of that ``triangle'' which
is linked to the number of extremities. But in a 3D space we can imagine
a ``lateral pyramidal'' growth from the branch $B$. This would
presumably give $\beta\simeq2$ using a similar reasoning. We can
imagine $\beta<1$ when the branch do not expand in the $Y$-direction
like the triangular shape, but instead remain approximatively confined
on a line. On the other hand, $\beta>3$ is hardly possible.

Replacing $V_{B}$ by $d^{\beta}$ in Eq.~\ref{eq:5}: $S\propto d^{2-\beta(\gamma-1)}$.
Because the threshold $S_{b}$ is a constant, when $S(d)$ is a decreasing
function it should mean structures that grew self-similarly and laterally
would not break, though no such protection exists once the self-similarity
ends and mechanical failures could also still happen to the ancestor-branches
supporting the structure. On the other hand, if $S(d)$ is an increasing
function, it will limit the size of such structure: indeed with $S(d)$
increasing as we go farther from the extremities $S(d)$ would increase
and the threshold $S_{b}$ would be reached resulting in a branch
more or less close to the base of the structure falling first. In
Sect.~\ref{subsec:Gravitational-loads}, $\gamma=3/2$ and we established
$\beta<3$ resulting in an increasing function.

Now we have another way to investigate the impact of gravitational
loads: changing $\gamma$. For example, for $\gamma>2$ would mean
$S(d)$ is decreasing even for $\beta=2$ which we argued it represents
a lateral pyramidal growth. And $\gamma>3$ guarantees even $\beta=1$
structures. However performing simulations with varying $\gamma$
around 2 and 3 do not result in any transition between radical behavior
in growth, the trees obtained are similar to what we described so
far. It could either be imputed to a weakness in our theory on $S(d)$
or the system is resilient to perturbation and factors that do not
directly affect redistribution schemes which would be consistent with
the observations we made the previous Sections. Some caveats: we are
not saying $\gamma$ did not have effect on growth, in fact as we
increase $\gamma$ occurrences where all branches up to the base of
trunk suddenly mechanically fail as described in Sect.~\ref{subsec:Gravitational-loads}
become more scarce. But it simply means the trees are now similar
to the ones in Sect.~\ref{subsec:Children-generated-in} (Fig.~\ref{fig:Fig5})
with no gravitational load: the tree does not seem to have more tendency
to grow straight.

\section{Conclusion\label{sec:Conclusion}}

We studied models combining network growth, source-sink and flow paradigm
and local resource allocation that were bio-inspired from trees. After
a brief summary of the previous paper\cite{Bui2019}, we try to understand
how resource allocation drives and controls the growth of the tree
as well as examine how the system responds to the addition of new
factors. More specifically, we test two resource allocation schemes:
the ``reward'' scheme which preferentially allocates resource to
more productive and wealthy nodes, and the ``maintenance'' scheme
which allocates according to the maintenance cost of each node. Both
schemes are parametrized by the exponent $a$ for ``reward'' and
$b$ for ``maintenance'' determining how much the allocation focuses
on reward or on maintenance (higher values of $a$ means a higher
focus on the productive nodes). As for the additional factors we submitted
the system to, the first one directly concerns resource allocation:
a small part of the resource is allocated to some nodes and we examine
the system reacts to it when using either ``reward'' or ``maintenance''
scheme. A different factor looked at was spatial embedding which limits
node creation by forbidding two nodes to occupy a same spatial location.
Finally, bio-inspired factors were added on top of the spatial embedding
such as gravitational loads created by the weight of the branches
themselves or light interception whereby leaves in the shadow of other
leaves or branches can not produce resource. We show the way the tree
grows is strongly driven by the resource allocation schemes once a
factor creates asymmetries. And such asymmetries can arise from a
simple embedding in space. On the other hand, the system is resilient
to further perturbations and factors when they do not directly act
on resource allocation schemes: the system is not affected by random
variables, changes in the values of parameters appears to affect continuously
and predictably the growth, and only light interception had an effect
on the shape. So when it comes to the shape of the crown, only few
shapes emerged and are attached to the way resource was allocated.
This may be interpreted as topology dominating geometry in this kind
of system as the only purpose our attempt to geometrize served was
to allow asymmetries.

Further investigations would be best focused on resource allocation
only. For example, the hierarchy created from the ``flux down''
- ``flux up'' scheme we used could be replaced with a scheme where
nodes would freely exchange to their neighbor. Moreover in the allocation
schemes we used all nodes prioritized either ``reward'' or ``maintenance''
the same way, but an alternative analysis could consist in allowing
different personalities within the same network.

\section*{Authors contribution statement}

O. Bui conducted all calculations made in the paper, X. Leoncini devised
the initial model and wrote the first version of the code for the
initial model introduced in \cite{Bui2019} and studied in the present
paper. Both authors discussed the research at its various stages,
and both contributed to the writing of the manuscript.


\end{document}